\documentclass[prl,aps,twocolumn,showpacs,floatfix]{revtex4-1}

\usepackage{epsfig}
\usepackage{amsmath}
\usepackage{amssymb}
\usepackage{amsfonts}
\usepackage{graphicx}
\usepackage{epsfig}
\usepackage{epstopdf}
\usepackage{bm}
\begin{document}

\title{Strain-driven nematicity of the odd-parity superconductivity in  Sr$_x$Bi$_2$Se$_3$.}

\author{A.~Yu.~Kuntsevich$^{a,b}$, M.A. Bryzgalov$^{a}$, R.S. Akzyanov$^{a,c,d}$, V. P. Martovitskii$^a$,  A.L. Rakhmanov$^{a,c,d,e}$, Yu. G. Selivanov$^a$ }
\affiliation{$^a$P.N. Lebedev Physical Institute of the RAS, 119991 Moscow, Russia}
\affiliation{$^b$National Research University Higher School of Economics, Moscow 101000, Russia}
\affiliation{$^c$Dukhov Research Institute of Automatics, Moscow, 127055 Russia}
\affiliation{$^d$Institute for Theoretical and Applied Electrodynamics, Russian
    Academy of Sciences, Moscow, 125412 Russia}
\affiliation{$^e$Moscow Institute of Physics and Technology, Dolgoprudny, Moscow Region, 141700, Russia}

\begin{abstract}
    In this letter, we present a novel experimental evidence for the odd-parity nematic superconductivity in high-quality single crystals of doped topological insulator Sr$_x$Bi$_2$Se$_3$. The X-ray diffraction shows that the grown single crystals are either weakly stretched or compressed uniaxially in the basal plane along one of the crystal axis. We show that in the superconducting state, the upper critical magnetic field $H_{c2}$ has a two-fold rotational symmetry and depends on the sign of the strain: in the stretched samples the maximum of $H_{c2}$ is achieved when the in-basal-plane magnetic field is transverse to the strain axis, while in the compressed samples this maximum is observed when the field is along the strain direction. This result is naturally explained within a framework of the odd-parity nematic superconductivity coupled to the strain. Magnetoresistance in the normal state is independent of the current direction and also has a two-fold rotational symmetry that demonstrates the nematicity of the electronic system in the normal state.
\end{abstract}

\maketitle
{\it Introduction.---} A specific crystal structure and strong spin-orbit coupling in the topological insulators (TIs) give rise to an unusual (topological) superconductivity in these materials~\cite{Fu2010}. A number of intriguing properties and, in particular, possible existence of Majorana fermions~\cite{Fu2008,Potter2013,Sarma2015,Elliott2015,Akzyanov2015,Akzyanov2016} attract a great attention to the superconductivity in the TIs~\cite{Sato2017}. A distinctive property of the TI is the existence of robust surface states with the Dirac spectrum~\cite{Zhang2009,Hasan2010}, and 2D topological superconductivity can be induced at the TI -- usual superconductor interface due to the proximity effect~\cite{Fu2008,Sau2010}. A bulk topological superconductivity with $T_c$ about 3~K was observed in the doped 3D TIs of the bismuth selenide family A$_x$Bi$_2$Se$_3$ (where A=Cu,Sr, and Nb)~\cite{Wray2010,Hor2010,Sasaki2011,Matano2016,Shen2017,Du2017,Pan2016, Kuntsevich2018,Tao2018,Yonezawa2016,Willa2018,Asaba2017,Smylie2018,Andersen2018,Chen2018,Cho2019,Sun2019}.

Knight shift below $T_c$ in  Cu$_x$Bi$_2$Se$_3$\cite{Matano2016} indicates nonzero-spin pairing in these materials. The superconducting order parameter in this system has a two-fold anisotropy which breaks the 3-fold symmetry of the crystals~\cite{Hor2010}. This rotational symmetry breaking is confirmed by transport~\cite{Pan2016, Shen2017, Du2017, Kuntsevich2018}, heat capacity~\cite{Yonezawa2016,Willa2018,Sun2019}, and magnetic measurements~\cite{Asaba2017,Smylie2018,Shen2017}. It was also demonstrated in the STM imaging as an observation of oval shape of the Abrikosov vertices~\cite{Tao2018}. This means the existence of the superconductivity with a nematic order parameter~\cite{Fu2010,Fu2014,Yonezawa2018}. {Note that the nematicity in the superconducting state is one of the recent central issues in condensed matter physics. This intriguing property is observed not only in TIs but also in different important materials (see, e.g, Refs.~\onlinecite{Li2019,le2019evidence,Andersen2018}).}

Among the irreducible representations of the $D_{3d}$ point group (inherent to A$_x$Bi$_2$Se$_3$), only multidimensional $E_g$ and $E_u$ representations of the superconducting order parameter are compatible with the experimental observations \cite{Fu2014}. These symmetries correspond to a nematic superconductivity with even ($E_g$) or odd ($E_u$) parity. The nodeless superconducting gap observed in the STM studies~\cite{Tao2018} and the specific heat data~\cite{Kriener2011,Yonezawa2016} are not compatible with $E_g$ pairing. Therefore, there exists an indication that the superconducting pairing in the A$_x$Bi$_2$Se$_3$ corresponds to the $E_u$ symmetry. The odd-parity nematic order parameter is a two-component vector having a definite direction~\cite{Fu2014, Venderbos2016}. There is a discrepancy in the experimental data about orientation of the nematicity director. In Refs.~\cite{Matano2016,Tao2018,Andersen2018} it has been argued that it is aligned along the main crystal axis, while the authors of Refs.~\cite{Yonezawa2016,Chen2018} concluded that it is perpendicular to that direction. An open question is also whether the rotational symmetry breaking is an externally induced (say, by an applied strain) or this symmetry breaking is spontaneous. Due to its vector nature, the nematic order parameter is non-trivially coupled with the strain~\cite{Venderbos2016}. It has been predicted that the nematicity is governed by the direction and sign of the strain. Therefore, the identification of the relationship between the nematicity direction and the strain is crucial for establishing the nature of the nematic superconductivity. {In Ref.~\onlinecite{Kuntsevich2018} an attempt to relate the nematicity in Sr$_x$Bi$_2$Se$_3$ and the crystal distortion was performed. However, the samples used in that paper were multi-block and the observed nematicity could be attributed to either an anisotropic arrangement of the crystal blocks or to the internal strain in the blocks. As
a result, a question remains open, whether the observed nematicity is a plain electromagnetic effect or it arises due to non-trivial electronic properties of the system.}

\begin{figure*}[t]
    \includegraphics[width=16cm]{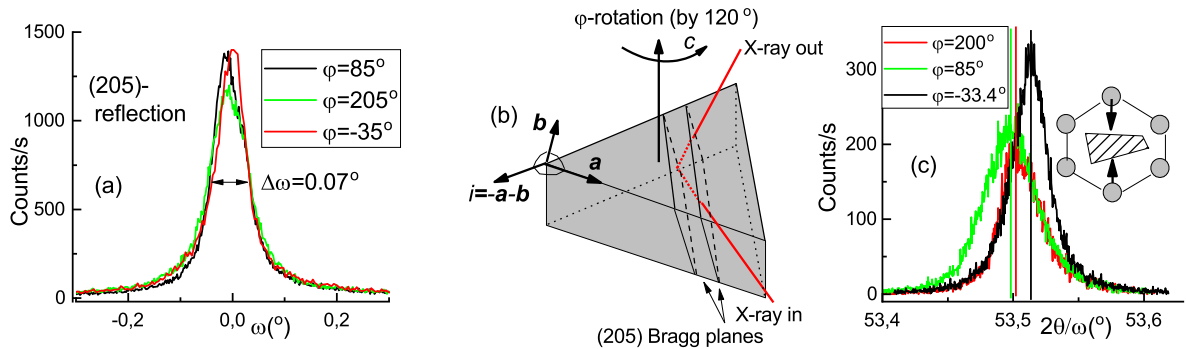}
    \caption{Structural studies of the samples. (a) Perfectly narrow and azimuthal angle-independent rocking curve for a single block sample taken at (205) XRD reflection. (b) Schematics of XRD experiment. Triangular shape of the prism corresponds the crystal symmetry. The X-ray enters and leaves the crystal through the top surface. The path of X-ray inside the crystal is shown by the dotted line.  (c) $2\theta/\omega$ scans at (205) reflection allowing to determine the direction of structural distortion (compression) with respect to the sample crystal axes (shown schematically in the inset). }
    \label{XRD}
\end{figure*}

{Here we report experiments performed on single-domain Sr$_x$Bi$_2$Se$_3$ samples. The results definitely show that the crystal deformation} (small in-basal-plane sample shortening or elongation) drives the direction of the superconducting order nematicity in Sr$_x$Bi$_2$Se$_3$. In particular, the upper critical {$c$-plane} magnetic field $H_{c2}$ depends on the sign of the strain: it has a maximum in the stretched samples when the applied magnetic field is transverse to the strain axis, while in the compressed samples this maximum is observed when the field is along the strain direction. These results were obtained on a number of high-quality Sr$_x$Bi$_2$Se$_3$ single crystals. Both value and direction of the deformation of the crystal structure was measured using high-resolution X-ray diffraction (XRD). The nematicity (that is, the anisotropy of $H_{c2}$) is suppressed significantly in the samples with low distortions. These results brilliantly confirm the existence of the multi-component odd-parity superconductivity in Sr$_x$Bi$_2$Se$_3$ and suggest the possibility to tune the superconducting properties of A$_x$Bi$_2$Se$_3$ by application of the uni-axial strain. {The second important result of the present study is that the crystal lattice distortion affects also the normal state properties far above $T_c$: similarly to superconductivity, the normal state {$c$-plane} magnetoresistance (MR) has two-fold rotational symmetry, linked to the crystal axes, and} independent of the angle between the magnetic field and the transport current, which demonstrates the nematicity of the electron system in the normal state. The MR is positive.

{\it Samples and XRD studies.---} We grew a series of Sr$_x$Bi$_2$Se$_3$ samples with nominal strontium content $x=0.1-- 0.2$  using a modified Bridgeman method (for details see Ref.~\cite{Kuntsevich2018}){Similar results were also obtained on Cu-co-doped Sr$_x$Bi$_2$Se$_3$ crystal\cite{Kuntsevich2019}. All} crystals consisted of blocks width a typical misorientation $\lesssim 1^\circ$. A common XRD techniques like powder~\cite{Liu2015,Pan2016} or Laue diffraction~\cite{Pan2016,Du2017,Sun2019} could not resolve this block structure and reveal fine features of each block. We  split the grown crystals into smaller pieces with typical dimensions {typically} not exceeding 1$\times$1$\times$0.1~mm$^3$ and hunt for single-block samples.

To characterize the obtained samples, we used XRD rocking curves, straightforwardly indicating a degree of misorientation of the crystal blocks. We selected the samples with narrow rocking curves. { This procedure is described in Section ``On the selection of single-block Sr$_x$Bi$_2$Se$_3$ single crystals'' of \cite{Supl} and also in Ref. \cite{Bond}.} An example of the remarkably narrow rocking curve for one of the selected samples is shown in Fig.~\ref{XRD}(a) for three different azimuthal angles $\phi$ at (2~0~5) reflection. Here we use the Miller indices in hexagonal notations (first two indices are in-basal-plane and third one is transverse to the plane, as described in { Section ``Definition of crystal planes in hegagonal notations'' of \cite{Supl}}). The widths of these curves are about 0.07$^\circ$; they are nearly independent of $\phi$ and the choice of a particular reflection. 

{We should note here, that  penetration depth of Cu $K_{\alpha}$ X-rays to Bi$_2$Se$_3$ is approximately 10 $\mu$m, whereas our samples have typically $50$–$80$ $\mu$m thickness. In order to ensure the single-crystallinity, we recorded rocking curves from both sides of the samples. Moreover, to exclude a possibility of twinning plane formation inside the crystals we measured $\phi$-scans on asymmetric reflections from also from both sides (see the last section of {\cite{Supl} with references \cite{Sungawa,Volosheniuk2019} used therein} for details).}

We can state that we deal with single crystals, uniform and free from grain boundaries. The samples used in the previous studies (see,e.g.,~\cite{Kuntsevich2018,Smylie2018}) have typical  rocking curve widths about 0.5$^\circ$, indicating  a multi-block structure. All XRD studies were performed using Panalytical MRD Extended diffractometer with a hybrid monochromator, that is a combination of a parabolic mirror and a single crystal 2 Ge(220) monochromator.

 \begin{figure*}
    \includegraphics[width=16cm]{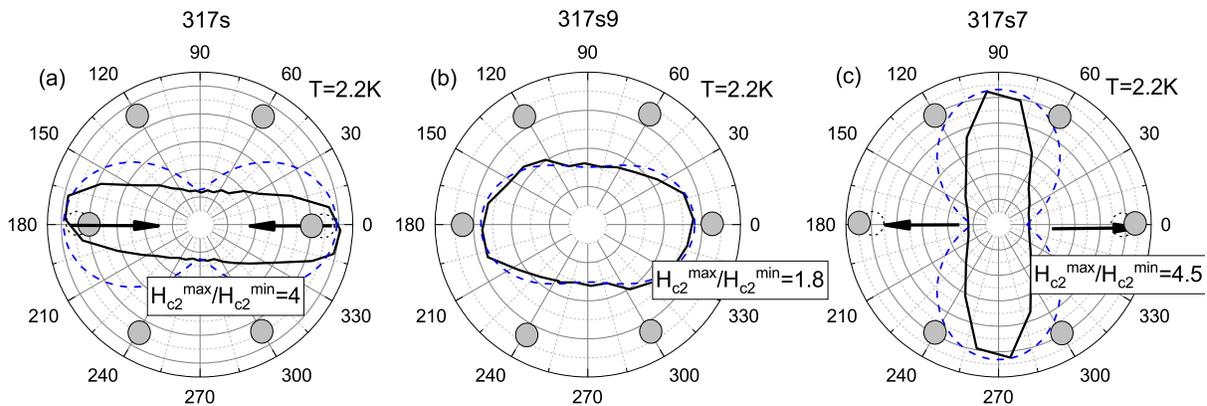}
    \caption{Dependence of $H_{c2}$ on the angle $\phi$ between the $c$-plain magnetic field and the strain axis for three samples: (a) compressed, (b) non-deformed, and (c) stretched; $T=2.2$~K. The directions of the strain axes are indicated by arrows. Dashed lines show the theoretical fit, Eq.~(\ref{Hc2}).
    }
    \label{supercond}
\end{figure*}

In order to find the in-basal-plane distortions in the pre-selected single-block samples we used high-resolution $2\theta/\omega$ XRD scans of asymmetric (205) reflection. For these precise measurements we used triple crystal-analyzer 3 Ge(220). The schematics of high-resolution XRD measurements is shown in Fig.~\ref{XRD}(b). The angle between Bragg planes and the basal plane is 72.7$^\circ$, therefore the (205)-XRD reflection is sensitive mostly to the lattice parameter $a$. The $2\theta/\omega$ position of the XRD peaks does not change when turning 120$^\circ$ around the $c$-axis if the crystal hass a trigonal symmetry. However, eight of nine selected samples were either stretched or compressed along one of the crystal axis and only one sample was non-deformed within available precision of the measurements. In the distorted samples the XRD peaks are slightly shifted on 120$^\circ$ rotation. An example of such shift in the peaks' position is illustrated in Fig.\ref{XRD}(c) for a compressed sample. The value of the shift allows us to estimate the strain. A characteristic deformation $\delta a/a \equiv \varepsilon_{xx}$ in our samples was about $\pm$~0.02\%. For each sample the direction and sign of the strain were determined before the transport measurements.

{\it Transport measurements, $H_{c2}$.---}
 The crystals were placed on a rotating platform of the cryomagnetic system (PPMS-9 and Cryogenics CFMS-16). From 4 to 7 contacts were attached to each sample. The dependencies of the sample resistance on the value and direction of the {$c$-plane} magnetic field were measured in the temperature range from $T=1.5$ to 300~K and in the magnetic field $B$ up to 15~T. Both XRD and transport {results} for nine samples are summarized in {Section ``Summary of structural and transport data for Sr$_x$Bi$_2$Se$_3$ single crystals'' of \cite{Supl}}.

The main results of our study are presented in Fig.~\ref{supercond} for three samples: compressed, non-deformed (within the experimental accuracy), and stretched, panels 2(a-c), respectively. These polar plots show the dependencies of the upper critical field $H_{c2}$ on the angle $\phi$ between the strain axis and the {$c$-plane} applied magnetic field $\bf{B}$ at $T=2.2$~K. The field $H_{c2}$ was determined  as a value of the field at which the resistance of the sample is half of the normal state one. In the compressed sample, the value $H_{c2}$ is maximum, when $\bf{B}$ is {\it parallel} to the direction of the compression, while in the stretched sample $H_{c2}$ has the maximum when $\bf{B}$ is {\it perpendicular} to the direction of the stretching. The anisotropy of $H_{c2}$ is about 4--4.5 for both deformed samples. The same relation between nematicity and the strain direction was observed in all samples (see {Section ``Summary of structural and transport data for Sr$_x$Bi$_2$Se$_3$ single crystals'' of \cite{Supl}}).

In the non-deformed at room temperature sample, Fig.~2(b), the nematicity in the superconducting state is also observed, however, with a significantly less anisotropy, $H_{c2}^{\text{max}}/H_{c2}^{\text {min}}\approx 1.8$. Still, the critical field anisotropy is noticeable. We can suggest that this is due to either spontaneous breaking of the rotational symmetry that occurs in the normal state near $T_c$~\cite{Yonezawa2018} or due to a limited experimental accuracy in measurement of $2\theta/\omega$, that is about $0.003^\circ$ {(see also discussion after Eq.~(\ref{Hc2}))}. Note also that the strain may vary with temperature. Further studies are necessary to clarify this issue.

The experimental results can be naturally explained within the framework of the Ginzburg-Landau (GL) theory for a superconductor with the order parameter ${\mathbf\eta}=(\eta_1,\eta_2)$ related to nematic odd-parity order parameter as ${\mathbf Q}=(|\eta_1|^2-|\eta_2|^2,\eta_1\eta_2^*+\eta_1^*\eta_2)$~\cite{Fu2014,Venderbos2016} (see also {Section ``Ginzburg-Landau theory of odd-parity nematic superconductivity'' of \cite{Supl}}). In the absence of the strain, all possible directions of the nematic director are equally favourable and the value of $H_{c2}$ is independent of the applied field direction. The strain generates a rotational symmetry breaking term in the GL free energy functional $f_{SB}$. In the case of the uniaxial deformation along the $x$-axis, $\varepsilon_{xx}$, this term is $f_{SB}= g\varepsilon_{xx}(|\eta_1|^2-|\eta_2|^2)$, where $g$ is a coupling constant. Thus, $f_{SB}$ depends not only on the strain value but also on its sign, and the nematicity director ${\mathbf n}$ changes its direction by $90^\circ$ with change of the sign of $\varepsilon_{xx}$. For example, if $g>0$, then, ${\mathbf
 n} = (0,1)$ for the stretched sample, $\varepsilon_{xx}>0$, and ${\mathbf n} = (1,0)$ for the compressed sample, $\varepsilon_{xx}<0$. As a result, $H_{c2}$ becomes a function of the the angle $\phi$ between the strain axis and the magnetic field.
 
\begin{figure*}
    \includegraphics[width=16cm]{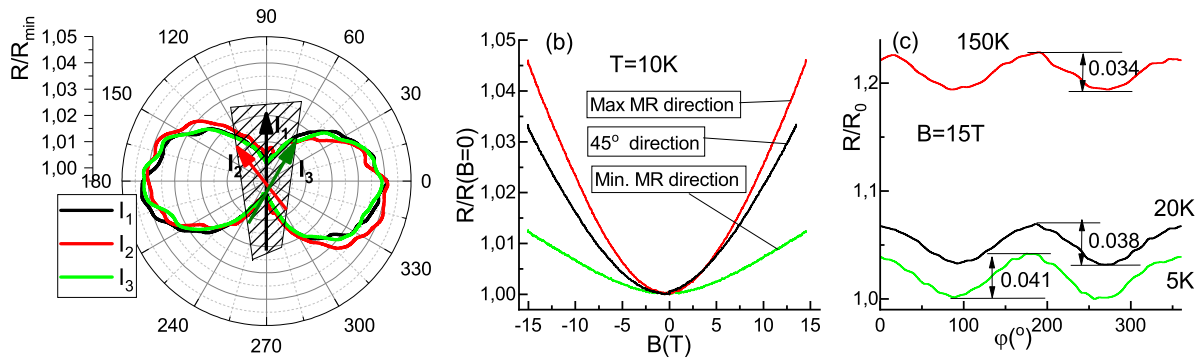}
    \caption{ Normal-state MR for the compressed sample no.~{317s} (see Fig.~\ref{supercond}(a)). (a) Polar plot of the normalized in-basal-plane MR $R/R_{\text {min}}$ as a function of the angle $\phi$ between the magnetic field and the strain axis for three different transport current directions $I_{1,2,3}$ in the same sample. The current directions are indicated by arrows, $R_{\text{min}}$ corresponds to the minimal resistance. (b) Dependence of the {$c$-plane} MR on the value of the magnetic field for different field orientations: red line corresponds to maximal MR direction, black line corresponds to the one shifted by the angle $45^\circ$, green line corresponds to the angle $90^\circ$ (minimal MR direction). The dependence is close to $\propto B^2$ in all cases. (c) Angular dependence of the normalized {$c$-plane} MR in the same sample taken at different temperatures. {The arrows indicate the amplitude of the effect.}}
    \label{normalstate}
\end{figure*}

{The strain of the sample
might modify the symmetry of the superconductivity due to a change of the crystal symmetry. For a weak strain, following Ref.~ \onlinecite{Venderbos2016}, we set the symmetry of the order parameter unperturbed, whereas the symmetry of $H_{c2}$ may be different. Then, we use expression for $H_{c2}(\varepsilon_{xx},\phi)$ from Ref.~\onlinecite{Venderbos2016} (see {Section ``Ginzburg-Landau theory of odd-parity nematic superconductivity'' of \cite{Supl}} for details). Solving it we obtain}
\begin{eqnarray}\label{Hc2}
&&\frac{H_{c2}(\varepsilon_{xx},\phi)}{H_{c2}(0)}=\\
\nonumber
&&\frac{1\!-\!a\epsilon\cos{2\phi}
+\sqrt{\left(1\!-\!a\epsilon\cos{2\phi}\right)^2\!-\!(1\!-\!a^2)\left(1\!-\!\epsilon^2\right)}}{1+|a|},
\end{eqnarray}
where $H_{c2}(0)$ is the upper critical field at $\varepsilon_{xx}=0$, $\epsilon=g\varepsilon_{xx}/|A|$, $a=J_4/2J_1$, and the values $A\propto T-T_c$ and $J_i$ are parameters of  the GL functional~\cite{Venderbos2016} defined in {\cite{Supl}}. The calculated dependence $H_{c2}(\phi)$ is shown in Fig.~\ref{supercond} at $a=0.7$ and $\epsilon=-0.6$ for the compressed sample and at $a=0.7$ and $\epsilon=0.64$ for the stretched sample. Thus, the GL theory for the odd-parity nematic superconductor allows us to describe qualitatively the experimental data. In particular, it properly reflects the fact that the nematicity direction changes by the angle $\pi/2$ with the change of the strain sign. Some discrepancy between calculated and measured dependence of $H_{c2}(\phi)$ may occur due to determination of $H_{c2}$ as a field at which the sample resistance is half of the normal value. The anisotropy of $H_{c2}(\phi)$ is also observed in the non-deformed sample, see Fig.~\ref{supercond}b. Its value is significantly smaller than that in the deformed crystals. The measured dependence $H_{c2}(\varepsilon_{xx},\phi)$ can be described by Eq.~(\ref{Hc2}) at $a=0.7$ and $\epsilon=-0.28$. { Note, the dependence $H_{c2}(\varepsilon_{xx})$ is quite non-linear and almost saturates for $\epsilon>0.6-0.7$.}  For a large strain, a different approach can be used for calculation of $H_{c2}$~\cite{Shen2017},  that even better describes our results for some samples. 

{\it Normal state magnetoresistance.---}  We performed a detailed investigation of the normal state MR in the {$c$-plane} magnetic field (see Fig.~\ref{normalstate} and {Section ``Summary of structural and transport data for Sr$_x$Bi$_2$Se$_3$ single crystals'' of \cite{Supl}} for more detail). Similar to $H_{c2}(\phi)$, the MR obeys the two-fold symmetry, and, surprisingly, the MR is independent of the current direction. This feature is illustrated in Fig.~\ref{normalstate}(a) where the normalized MR is shown as a function of $\phi$ for three different current paths in the same sample. {This observation indicates irrelevance of the Lorentz force to the normal state MR.}

In the previous studies the anisotropy of the MR was attributed either to the hexagonal warping of the Fermi surface~\cite{Akzyanov2018}, presence of the magnetic impurities~\cite{Chiba2017} or inversion-symmetry breaking near the sample interface~\cite{Taskin2017}. According to our experiments, the MR anisotropy is linked to some selected {$c$-plane} direction of the crystal. This direction is not connected unambiguously to the nematicity direction in the superconducting state. From sample to sample the maximal $H_{c2}$ and maximal MR directions may be parallel, perpendicular or makes an angle about $60^{\circ}$ to each other (see {\cite{Supl}}). This puzzling behaviour of the MR requires further investigation.

The measured {$c$-plane} MR is positive and its dependence on $B$ is almost quadratic, Fig.~\ref{normalstate}(b). The former result contradicts Refs.~\cite{Kuntsevich2018,Huang2018} for large multi-block samples, where a negative MR was observed. We attribute this contradiction to a contribution of the grain boundaries in the larger samples.

The value of the normal state anisotropic MR, $[R(\phi)-R_{\text{min}}]/R_{\text{min}}$, {weakly drops with temperature, Fig.~\ref{normalstate}(c). The two-fold anisotropy of the MR in a wide temperature range indicates that a nematic order exists in the normal state. It also means that there are no structural transitions during the cool down of the samples. The most probably, it is the distortion seen from the room-temperature XRD data that drives the superconducting nematicity.}

In {Section ``Magnetoresistance in a normal state generated by strain'' of \cite{Supl} (see also references \cite{fu2009,Proskurin,Akzyanov2018,Chiba2017,Taskin2017} used therein)} we present a theory of the strain-induced normal state MR based on the low-energy Hamiltonian of the TI. The theory catches the main features of the observed MR ({weak dependence on $T$, quadratic field dependence, and independence of the MR on current direction.} However, it can not explain the variation of the MR anisotropy axis from sample to sample.

Recently, there was a theoretical suggestion~\cite{Hecker2018} and experimental indications~\cite{Sun2019,Cho2019} for the vestigial order generated by the nematic superconducting fluctuations in doped Bi$_2$Se$_3$ slightly above $T_c$. This fascinating fluctuation effect is out of the scope of our paper and it does not contradict to our observations that small internal strain drives nematicity direction.

{\it Conclusions.---}  {We observe} a one-to-one correspondence between the structural distortion (small stretching or compression) and superconducting nematicity in the single crystals of doped 3D TI Sr$_x$Bi$_2$Se$_3$: in the stretched samples the maximum of $H_{c2}$ is achieved when the {$c$-plane} magnetic field is transverse to the strain axis and in the compressed crystals this maximum is observed when the field is along the strain direction. This finding is a {definite} experimental signature of the odd-parity $E_u$ nematic superconductivity, that is supported by speculations based on the two-component order parameter Ginzburg-Landau theory. The normal state magnetoresistance in the strained samples {also exhibits nematicity far above $T_c$. Thus, we reveal that a weak strain of Sr$_x$Bi$_2$Se$_3$ crystal produces a significant effect on electronic properties in a wide temperature range}.

\begin{acknowledgments}
The work is supported by the Russian Science Foundation (Grant No. 17-12-01544). RSA acknowledges partial support by the Basis foundation. The measurements were performed in P.N. Lebedev Physical Institute shared facility center. The authors are thankful to V.V. Kabanov, Ya. A. Gerasimenko, V.A. Prudkogliad for discussions.
\end{acknowledgments}

\bibliographystyle{apsrev4-1}

\clearpage

\begin{widetext}
{\bf{\Large Supplementary: Strain-driven nematicity of the odd-parity superconductivity in  Sr$_x$Bi$_2$Se$_3$.}}

\maketitle

\section{Ginzburg-Landau theory of odd-parity nematic superconductivity}

A weak strain significantly affects the observed superconducting properties of Sr$_x$Bi$_2$Se$_3$. 
According to the experimental results shown in Fig.~2 of the main text, different types of the $H_{c2}(\phi)$ anisotropy correspond to the different signs of the sample strain, compressed ($\varepsilon_{xx}<0$) and stretched ($\varepsilon_{xx}>0$). This feature is naturally explained in terms of the two-component nematic superconductivity with odd parity proposed in Refs.~\onlinecite{Fu2010,Fu2014,Venderbos2016}. The Ginzburg-Landau (GL) free energy of such a superconductor can be written in the form~\cite{Fu2014,Venderbos2016}
\begin{equation}\label{G_L}
f_{\textrm{GL}}=f_{\textrm{hom}}+f_D+f_{SB},
\end{equation}
where the GL free energy of the homogeneous superconducting state is
\begin{equation}\label{hom}
f_{\textrm{hom}}\!=\!A\!\left(|\eta_1|^2\!+\!|\eta_2\!|^2\right)\!+\!u_1\!\left(|\eta_1|^2\!+\!|\eta_2|^2\right)^2\!+\!u_2\!\left(\eta_1^2\!+\!\eta_2^2\right)^2,
\end{equation}
$\mathbf{\eta}=(\eta_1,\eta_2)$ is a two-component superconducting order parameter, $A\propto T-T_c$, $u_1>0$, and $u_2$ are phenomenological GL coefficients. The gradient part of the free energy is
\begin{eqnarray}\label{grad}
\nonumber
f_D&=&J_1(D_i\eta_\alpha)^*D_i\eta_\alpha+J_3(D_z\eta_\alpha)^*D_z\eta_\alpha+J_4\big[|D_x\eta_1|^2 \\
&+&|D_y\eta_2|^2\!-\!|D_x\eta_2|^2\!-\!|D_y\eta_1|^2\!+\!(D_x\eta_1)^*D_y\eta_2\\
\nonumber
&+&(D_y\eta_1)^*D_x\eta_2+(D_x\eta_2)^*D_y\eta_1+(D_y\eta_2)^*D_x\eta_1\big],
\end{eqnarray}
where the gauge-invariant gradient $D_i=-i\partial_i-(2e/c)A_i$, $\mathbf{A}$ is the electromagnetic vector potential, $J_i$ are the GL phenomenological coefficients, and summation over repeated indices $i=x,y$ and $\alpha=1,2$ is assumed. The GL theory with the free energy $f_{\textrm{GL}}=f_{\textrm{hom}}+f_D$ describes a superconductivity with a nematic order parameter $\mathbf{Q}=(|\eta_1|^2-|\eta_2|^2,\eta_1^*\eta_2+\eta_2^*\eta_1)$~\cite{Fu}. All possible directions of the nematicity $\mathbf{n}=(\cos \phi, \sin \phi)$ are equally favourable and the value of the upper critical field $H_{c2}$ is independent of the applied field direction with respect to the crystal axis, that is, independent of $\phi$. 

The strain of the sample generates a rotational symmetry breaking term $f_{SB}$. And this contribution can be written as~\cite{Fu1014,Venderbos2016}
\begin{equation}\label{f_SB}
f_{SB}=g\varepsilon_{xx}\left(|\eta_1|^2-|\eta_2|^2\right),
\end{equation}
if the strain is uniaxial, $\varepsilon_{xx}$. Here $g$ is the coupling constant. As a result, nematicity acquires a preferable direction.  Moreover, the value $H_{c2}(\phi)$ depends not only on the direction of the strain, but also on its sign. 

{Lowering the crystal symmetry due to distortions leads to the change of the symmetries of the superconducting order parameter. However, the crystal strain in our samples is tiny (about 0.02 \% and smaller). Thus, it is convenient to consider the symmetry class of the crystal without deformation and add deformation via the perturbation theory as it was done in Ref.~ \onlinecite{Fu2}. Following the latter work, we assume that the breaking of $C_3$ symmetry changes considerably only the kinetic part of the free energy. Thus, we expect that the symmetry of the order parameter would be the same for our strained and non-deformed samples, but the symmetry of the second critical field $H_{c2}$ may be different.}
The analytic equation for $H_{c2}(\phi)$ was found in Ref.~\onlinecite{Venderbos2016} [see Eq.~(15) from this reference]. That equation can be easily converted to a quadratic one and we obtain an explicit formula for $H_{c2}(\phi)$ in the form
\begin{eqnarray}\label{hc2}
\frac{H_{c2}(\varepsilon_{xx},\phi)}{H_{c2}(0)}=\frac{1\!-\!a\epsilon\cos{2\phi}
+\sqrt{\left(1\!-\!a\epsilon\cos{2\phi}\right)^2\!-\!(1\!-\!a^2)\left(1\!-\!\epsilon^2\right)}}{1+|a|},
\end{eqnarray}
where $H_{c2}(0)=|A|/\left[\sqrt{|J_1J_3|}(1-|a|)\right]$ is the upper critical field at $\varepsilon_{xx}=0$ (which is independent of $\phi$), $\epsilon=g\varepsilon_{xx}/|A|$, and $a=J_4/2J_1$. 

The dependence $H_{c2}(\phi)$ is shown in Fig.~S1%\ref{figHc2}
. The function $H_{c2}(\phi)$ has a two-fold symmetry. This function has maximums either at $\phi=0$ and $\pi$ or at $\phi=\pi/2$ and $3\pi/2$ (see Fig.~S1
%\ref{figHc2}
) depending on the sign of the product $ag\varepsilon_{xx}$. The sign of the strain is known from the experiment: $\varepsilon_{xx}<0$ for the compressed samples and $\varepsilon_{xx}>0$ for the stretched samples. The signs of $a$ and $g$ can not be extracted from a general symmetry consideration. However, based on the experimental data shown in Fig.~2 of the main text, we can conclude that for our crystals $ag>0$. In the main text, we assume for definiteness that both $a$ and $g$ are positive. 

The theory qualitatively describes the experiment as it follows from the results shown in Fig.~2 of the main text. Some difference between the theoretical predictions and measured behavior of $H_{c2}(\phi)$ can be attributed to the definition of this value from the experiment, which is rather voluntaristic. On the other hand, this discrepancy can occur due to neglecting of higher order terms in the GL free energy expansion Eqs.~(\ref{hom}) and/or (\ref{grad}).       

%In conclusion of this section we note that dimensionless parameter $\epsilon$, which characterizes the nematicity, increases as the temperature $T$ approaches the superconducting transition temperature $T_c$: $\epsilon\propto (T_c-T)^{-1}$. Therefore, the anisotropy of the upper critical field in the strained samples increases when $T\rightarrow T_c$.  

\begin{figure}[t!]
	\center
	\includegraphics[width=12 cm]{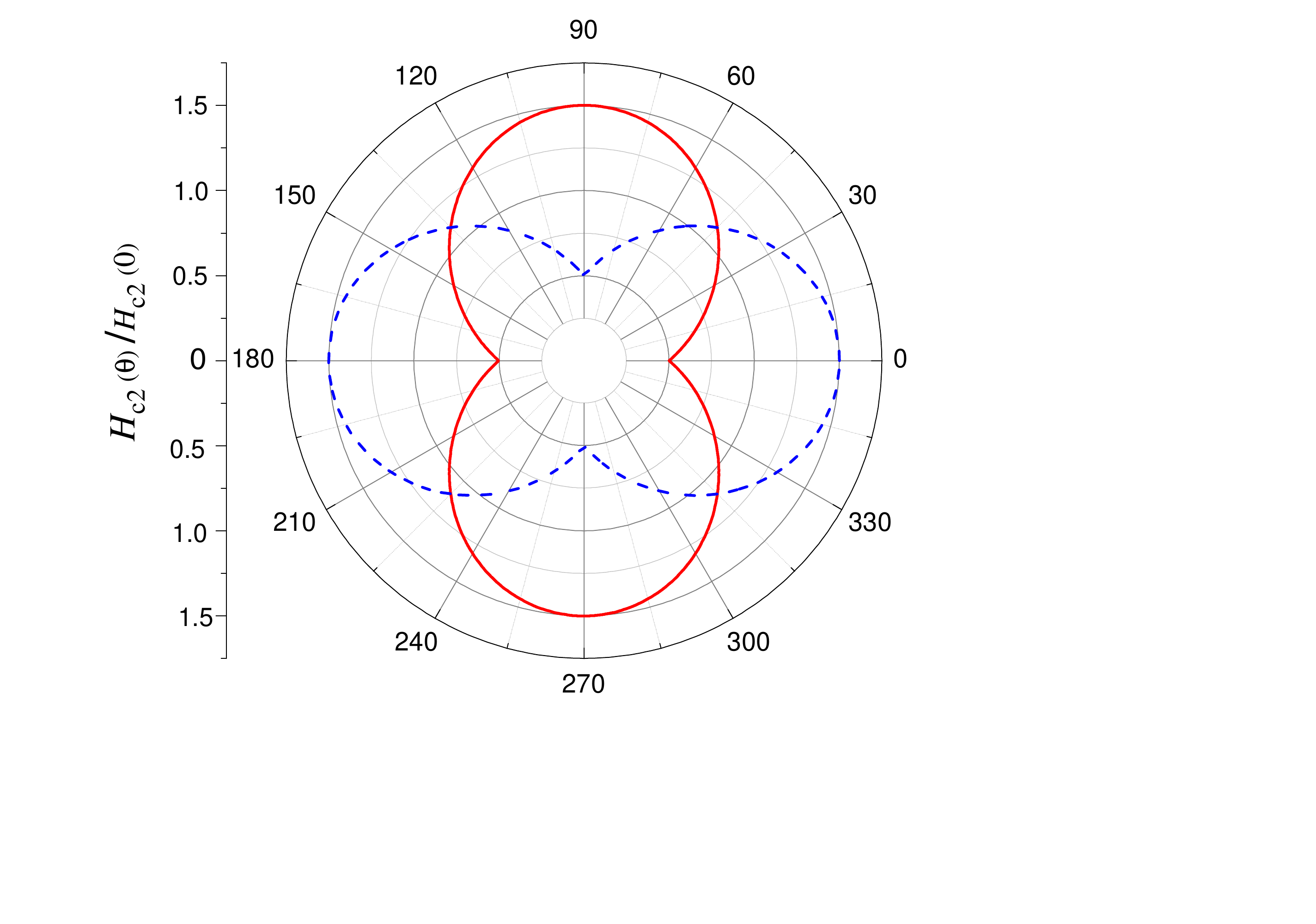}
	{\\Fig. S1. Dependence of $H_{c2}(\phi)/H_{c2}(0)$; $a=0.5$, $\epsilon=0.5$ (red solid line) and $\epsilon=-0.5$ (blue dash line). The anisotropy of the upper critical field is about 3 for these parameters. The minimum value of $H_{c2}(\phi)$ for the strained sample is half of $H_{c2}(0)$, while its maximum value is $1.5H_{c2}(0)$}.%\label{figHc2}
\end{figure}

\section{Magnetoresistance in a normal state generated by the strain}

Hamiltonian for the bulk quasi-two dimensional states in the TI Bi$_2$Se$_3$ obeys a $C_{3v}$ symmetry. That is, it is invariant with respect to a time-reversal symmetry with operator $T=i\sigma_yK$ (where $K$ means complex conjugation and $\sigma_i$ are the Pauli matrices acting in the spin space) and two crystal symmetries: mirror symmetry $x\rightarrow -x$ with operator $M=i\sigma_x$, and rotational symmetry $C_3=e^{-i\sigma_z\pi/3}$. Up to the second order in momentum $\mathbf{k}=(k_x,k_y)$, minimal Hamiltonian that obeys such symmetry can be written in the form~\cite{fu2009}
\begin{equation}\label{H0}
H_0=\mu+r(k_x^2+k_y^2) +\alpha_R(k_x\sigma_y-k_y\sigma_x),
\end{equation}
where $\mu$ is the chemical potential, $r=1/2m$ is the inverse mass term, and $\alpha_R$ is the Rashba coupling. Note, that this Hamiltonian is not necessary true Hamiltonian for the bulk states in the real topological insulator, but it is a minimal model that obeys given symmetries.

The strain $u_{xx}$ along the $x$ direction breaks $C_3$ symmetry but preserves the mirror $M$ and the time-reversal $T$ symmetries. In general, the strain brings anisotropy in the mass term $r$ and the Rashba coupling $\alpha_R$. If the strain is small, then, the corrections to these terms are small and can be neglected. Otherwise, the $C_3$ symmetry is broken and the additional term
\begin{equation}
H_s=\epsilon k_x \sigma_z
\end{equation}
is allowed in the Hamiltonian. It acts as a source of the anisotropic magnetoresistance (MR). 

We assume that the in-plane magnetic field $\mathbf{B}=(B\cos\phi,B\sin\phi)$ is small, $B \ll \mu$, and we can neglect its influence on the impurity-averaged scattering amplitude $\Gamma$. Strain axis $x$ corresponds to $\phi=0$. We calculate the longitudinal conductivity with the Bastin-Kubo formula, which can be written in the form~\cite{Proskurin}:
\begin{equation}
\sigma_{ii} = \frac{\sigma_0}{2}\sum_{\bf k} v_i (G^+-G^-)v_i(G^+-G^-),
\end{equation}
where $\sigma_0=e^2/{\hbar}$, $G^{\pm}=(H_0+H_s+\sigma_xB_x+\sigma_yB_y\pm i\Gamma)^{-1}$ are the impurity averaged advanced/retarted Green's functions, and $v_i=\partial (H_0+H_s)/\partial k_i$ is the velocity operator. We suppose that under experimental conditions $R_{ii}\simeq 1/\sigma_{ii}$. Then, MR is calculated as $MR=R(\phi)/R(0)$. The results of the calculations are shown in Fig.~S2%\ref{figMR}
. We observe two minimums (maximums) when the magnetic field is directed along (transverse) the strain axis. The rotational symmetry of the MR is independent of the current direction. The value of $MR$ is different for different current directions. However, this difference is quite small. This result completely coincides with the experiment, see  Fig.~3 of the main text. Resistance increases quadratic, as $B^2$, with the increase of the magnetic field, Fig.~S3
%\ref{b2}
. Such increase is largest if the magnetic field is transverse to the strain direction. The resistance practically independent of the magnetic field if the field is parallel to the strain axis. This result also coincides well with the experiment, Fig.~3 of the main text. We fit experimental data using reasonable parameters: $r\mu/\alpha_R^2=-2$, $r\Gamma/\alpha_R^2=0.1$, and $\epsilon/\alpha_R=0.6$, Fig.~S3
%\ref{b2}
. Some discrepancy between observed and calculated MR at $\phi=\pi/2$ can occur due to accuracy of the measurements of the angle $\phi$ in the experiment.   

\begin{figure}[t!]
	\center
	\includegraphics[width=8.5cm, height=6cm]{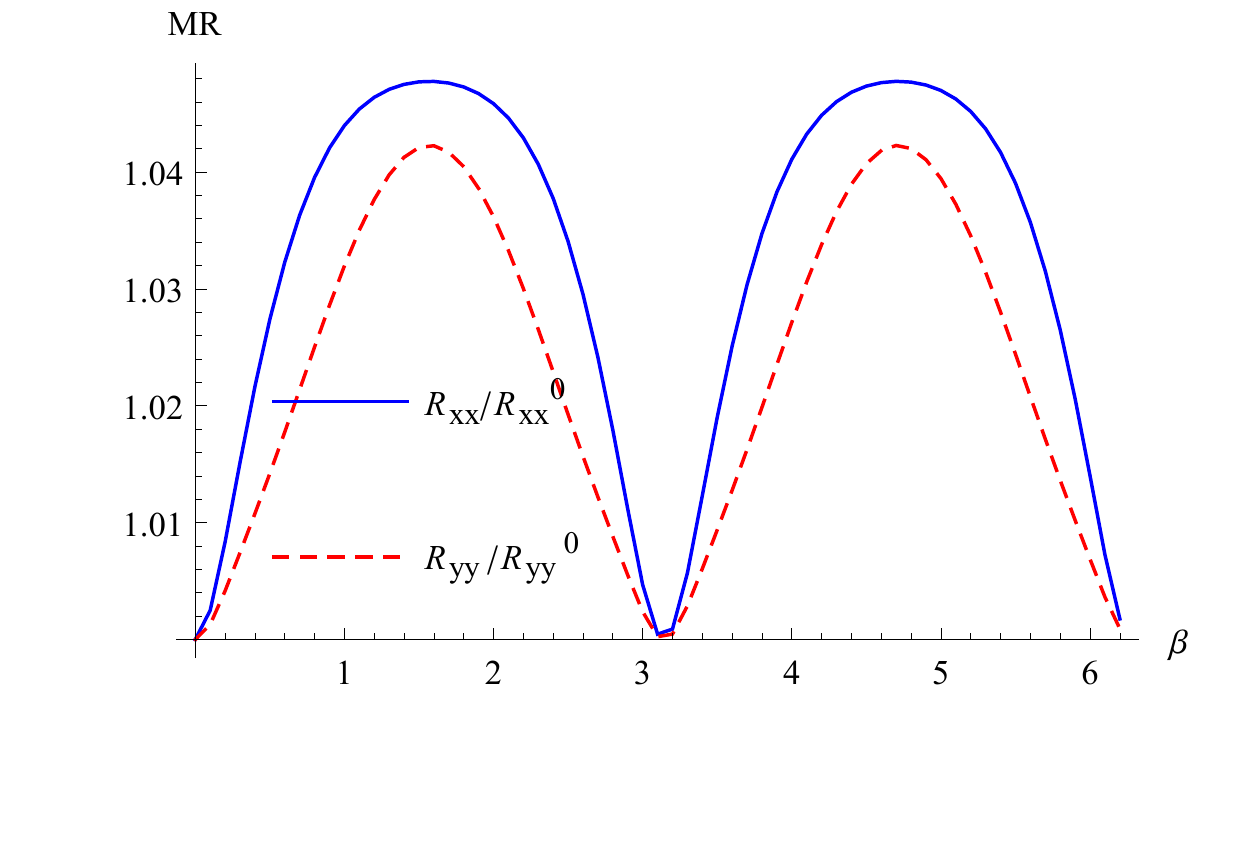}
	{\\
	Fig. S2. Magnetoresistance  $MR=R(\beta)/R(\beta=0)$ versus the angle $\phi$ between the magnetic field and the strain axis $x$. Blue line corresponds to $R_{xx}(\beta)/R_{xx}(0)$, that is the MR if current is parallel to the strain axis. Red line corresponds to the MR if current is transverse to the strain axis, $R_{yy}(\beta)/R_{yy}(0)$. }\label{figMR}
\end{figure}

\begin{figure}[t!]
	\center
	\includegraphics[width=8.5cm, height=6cm]{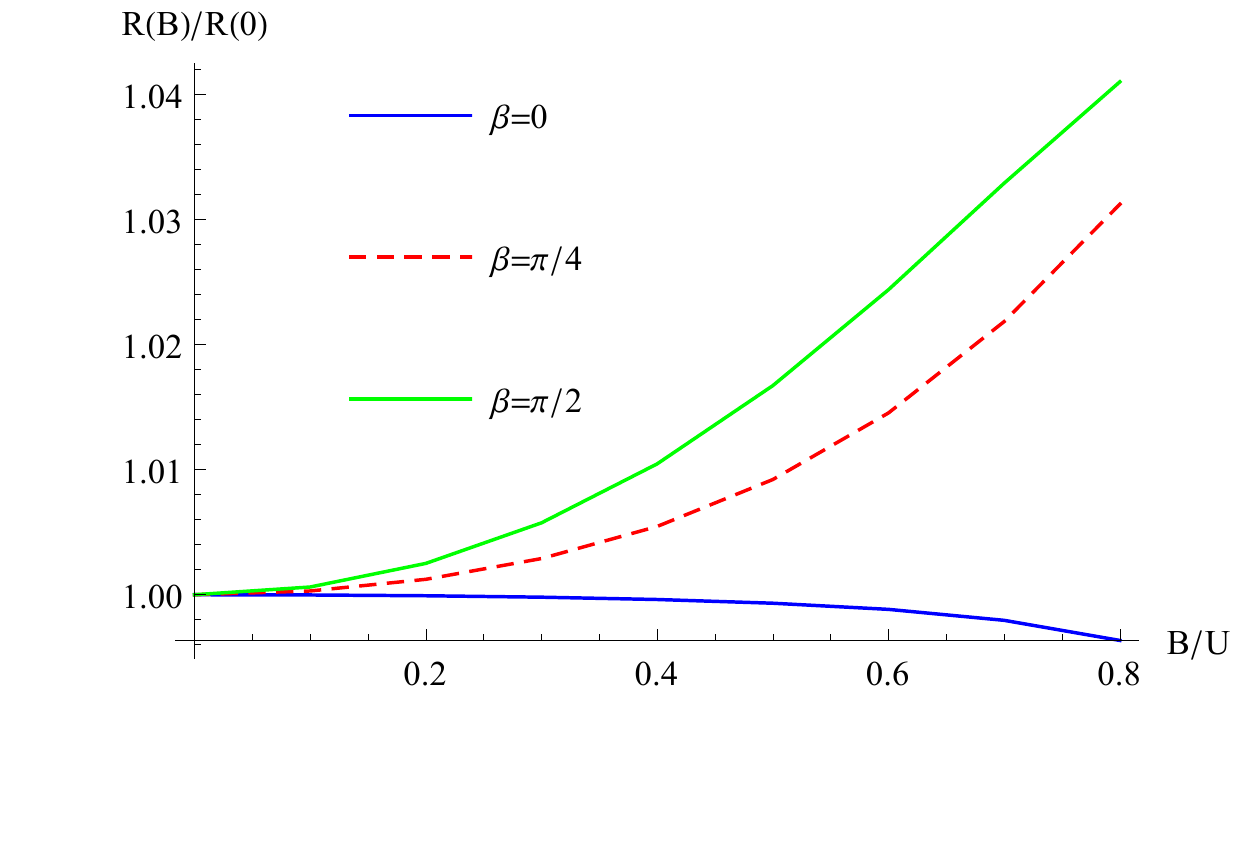}
	{\\Fig S3. Magnetoresistance $R_{xx}(B)/R_{xx}(B=0)$ as a function of the magnetic field $B$. Blue line corresponds to the case of the magnetic field aligned with the strain axis $\phi=0$, red line corresponds to the case when the magnetic field has an angle $\phi=\pi/4$ with the strain axis, green line corresponds the case when the magnetic field is perpendicular to the strain axis, $\phi=\pi/2$}.\label{b2}
\end{figure}

{The obtained results are valid if $B\ll u$. This means that the Zeeman field $B$ is much less than the Fermi energy $\mu$  counted from the bottom of the conductivity band. The ARPES data gives that $\mu\sim 100$meV~\cite{Liu2015}. For the magnetic field $B\sim 10$T and the characteristic g-factor $g=20$. Then, we have an estimate $B=g\mu_B H/2\sim 6$meV. Thus, the assumption $B\ll \mu$ is well satisfied.}

Different mechanisms could be a reason for the anisotropy of the MR in the TI placed in the in-plane magnetic field. Among the most frequently mentioned are hexagonal warping~\cite{Akzyanov2018} and the presence of the magnetic impurities with magnetization aligned along the applied magnetic field~\cite{Chiba2017,Taskin2017}. However, in these cases the anisotropy of the MR dependent strongly on the angle between the current and magnetic field, while in the presented model it is almost independent of the current direction. 

\newpage

\section{On the selection of single-block ${\rm Sr}_x{\rm Bi}_2{\rm Se}_3$ single crystals.}

Large single crystals, as pointed in the main text, consist of many blocks. In practice, to identify the presence of these blocks it does not matter whichever XRD reflection is used for rocking curve scans. Fig.%\ref{XRDSelection}
S4a shows a photo and a rocking curve of a large (1~cm linear size) multi-domain single crystal. Individual blocks are clearly seen in this scan (shown by the arrows). Total number of blocks ($\sim 10$) means that their linear dimensions are about 1 mm.
So, we split the crystal into pieces trying to find a single-block sample.
Fig.~%\ref{XRDSelection}
S4b shows an example of such single block crystal (mounted on a platform for transport measurements) and the corresponding ultra-narrow rocking curve.

\begin{figure}[h!]
\begin{minipage}{\textwidth}
  \includegraphics[width=0.8\textwidth]{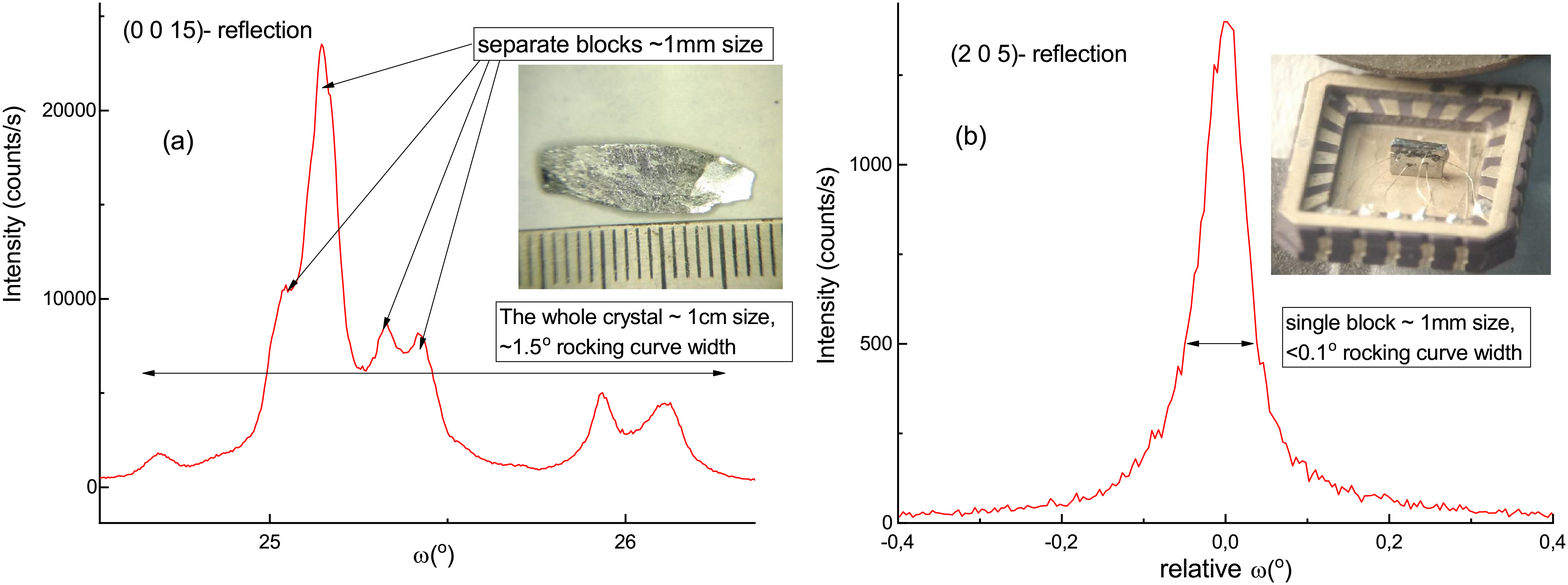}
  {\\ Fig. S4. Rocking curves and photos (in the inserts) of (a) large boule of Sr$_x$Bi$_2$Se$_3$ with approximate length 1.5~cm, and (b) single block crystal (length $\sim 1.3$ mm), glued on a Si wafer and mounted on a platform for transport measurements. Reflection indices are indicated in the panels.}
  \label{XRDSelection}
\end{minipage}
\end{figure}

{
\section{Definition of crystal planes in hexagonal notations.}
\begin{figure}[t!]
	\center
	\includegraphics[width=13 cm]{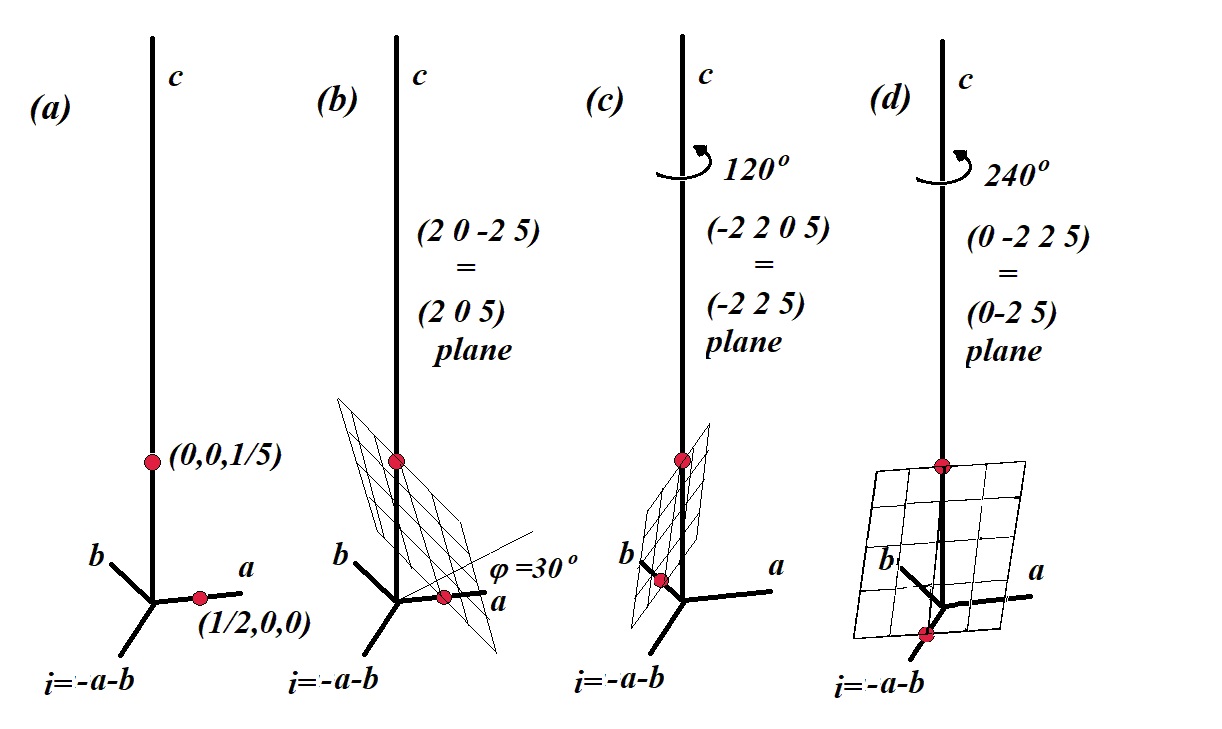}
	{\\ Fig.S23 Schematic reperesentation of the crystalline axes for Bi$_2$Se$_3$ in hexagonal notations. (a) The points used to plot (205) plane; (b) (205) plane and the corresponding $\phi=30^\circ$ value; (c-d) (-2 2 5) and (0 -2 5) planes, respectively.}\label{205scheme}
\end{figure}
Bismuth selenide belongs to R$\overline{3}$m symmetry group (trigonal singony), with the basis vectors forming the edges of rhombohedron. As the material is layered, and the physics inside layers and between them is usually considered separately,  in practice,  hexagonal notations are used for Miller indexes.

The hexagonal notations might be written in the 3-index form $(a~b~c)$ or 4-index form $(a~b~i~c)$. Here $c$ - is the out-of basal plane index, $a,b,i$ are the in-plane indexes, and the value of $i$ is equal to $-a-b$ by definition (correspondingly $b=-i-a$, and $a=-i-b$).

We show the crystalline directions (with the lengths proportional to those in Bi$_2$Se$_3$) in Fig.~S23%\ref{205scheme}
.

Now let us consider the orientation of the (2 0 5)-plane, that was the most explored throughout the paper. In order to define (2 0 5)-plane, we take the intervals 1/2 along $a$-axis, $1/5$ along $c$-axis. As the second index is 0, the plane is parallel to $b$-axis (see Fig.~S23%\ref{205scheme}
a-b). The value of $i$-index is automatically equal to -2.

Importantly, the perpendicular to this plane has the azimuthal angle $\phi=30^\circ$, because the plane is parallel to $b$ axis. 

Figures S23%\ref{205scheme}
b,c show the planes, obtained by rotation of the sample by 120$^\circ$ and 240$^\circ$ around $c$-axis, respectively. The in-basal plane indices ($a,b,i$) include one zero index, and, correspondingly, two indices with the opposite values, because $a+b+i=0$.
}

\newpage
\section{Summary of structural and transport data for ${\rm Sr}_x{\rm Bi}_2{\rm Se}_3$ single crystals}

All samples were preliminary selected by narrow rocking curves and further XRD-characterized{, as explained in the textbook \cite{Bond} }. They all demonstrated positive and weakly $T$-dependent magnetoresistance (independent of the current direction, as we explained in the main text) and nematic superconductivity (also independent of the transport current flow direction).
We summarize below XRD $2\theta/\omega$ scans and polar plots with magnetoresistance in both normal state (typically 5 or 20~K, and magnetic field 6-15 T as indicated) and in the resistive state (i.e at the boundary between superconducting and normal state). Red scale bar shows the scale of the normal state magnetoresistance. %8-like magnetoresistance in the resistive state is, apparently, perpendicular to maximal $H_{c2}$ direction.
For each sample we summarize the main nematic properties and give the nominal Sr content $x$. Note, that the real Sr content typically saturates at the level $\sim 0.06$ \cite{Liu2015}.

\begin{figure}[h!]
{\bf Sample S317s.}\\
The 317s sample ($x=0.15$) is compressed along one of the $c$-plane crystallographic directions (discussed also in the main text). {\bf Direction of the maximal $H_{c2}$ is parallel to the compression axis}. Direction of maximal normal magnetoresistance is rotated 60$^o$ from the compression axis. Anisotropy factor $H_{c2}^{max}/H_{c2}^{min}$ = 4. $\Delta a/a = 0.028\%$\\
\vspace{-0.1in}
\begin{minipage}{.49\textwidth}
  \includegraphics[width=6.5cm]{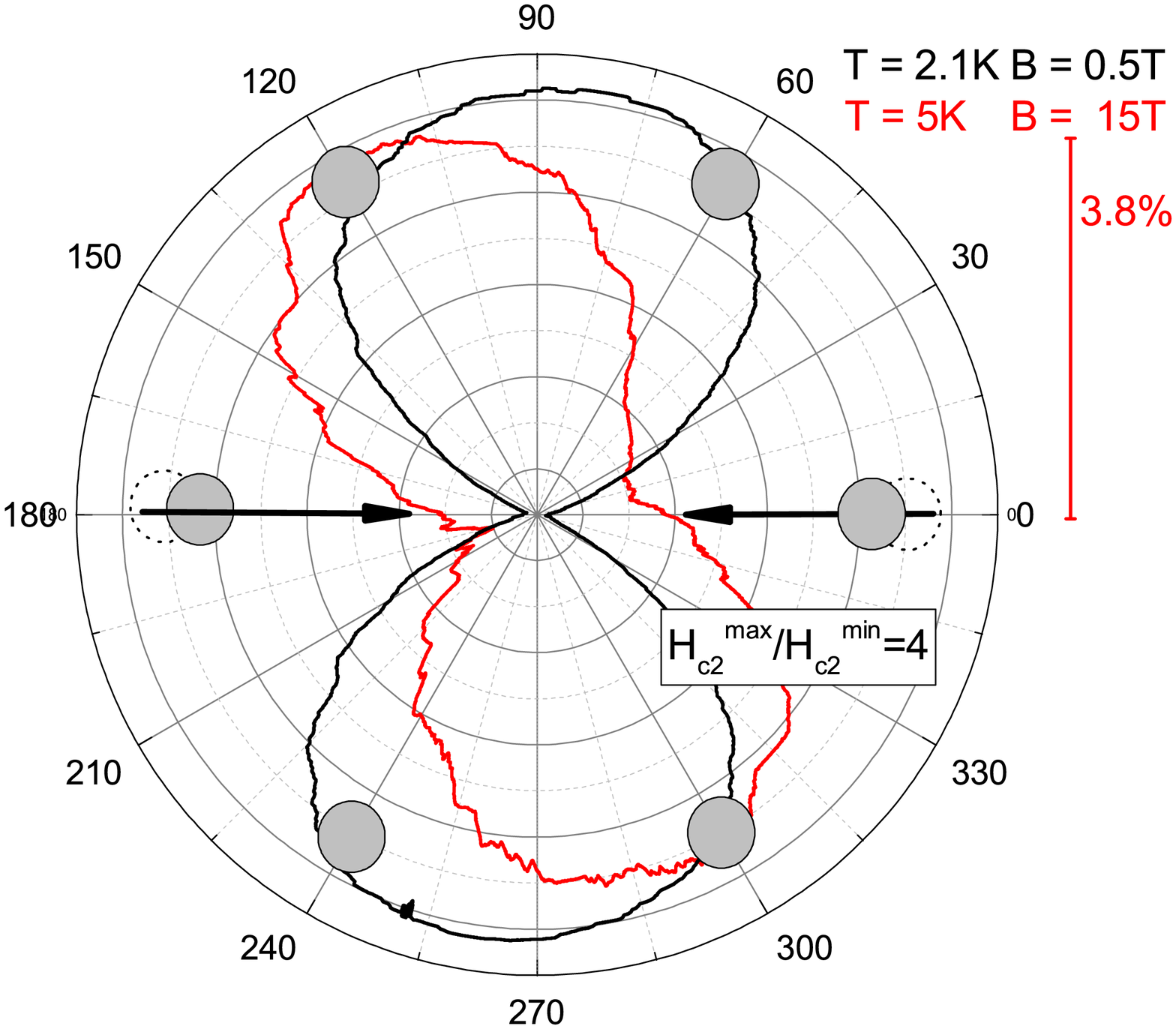}
\vspace{-0.1in}
  {\\ Fig.S5. Sample 317s. Magnetoresistance above and below $T_c$.}
  \label{317s}
\end{minipage}
\begin{minipage}{.49\textwidth}
  \includegraphics[width=8cm]{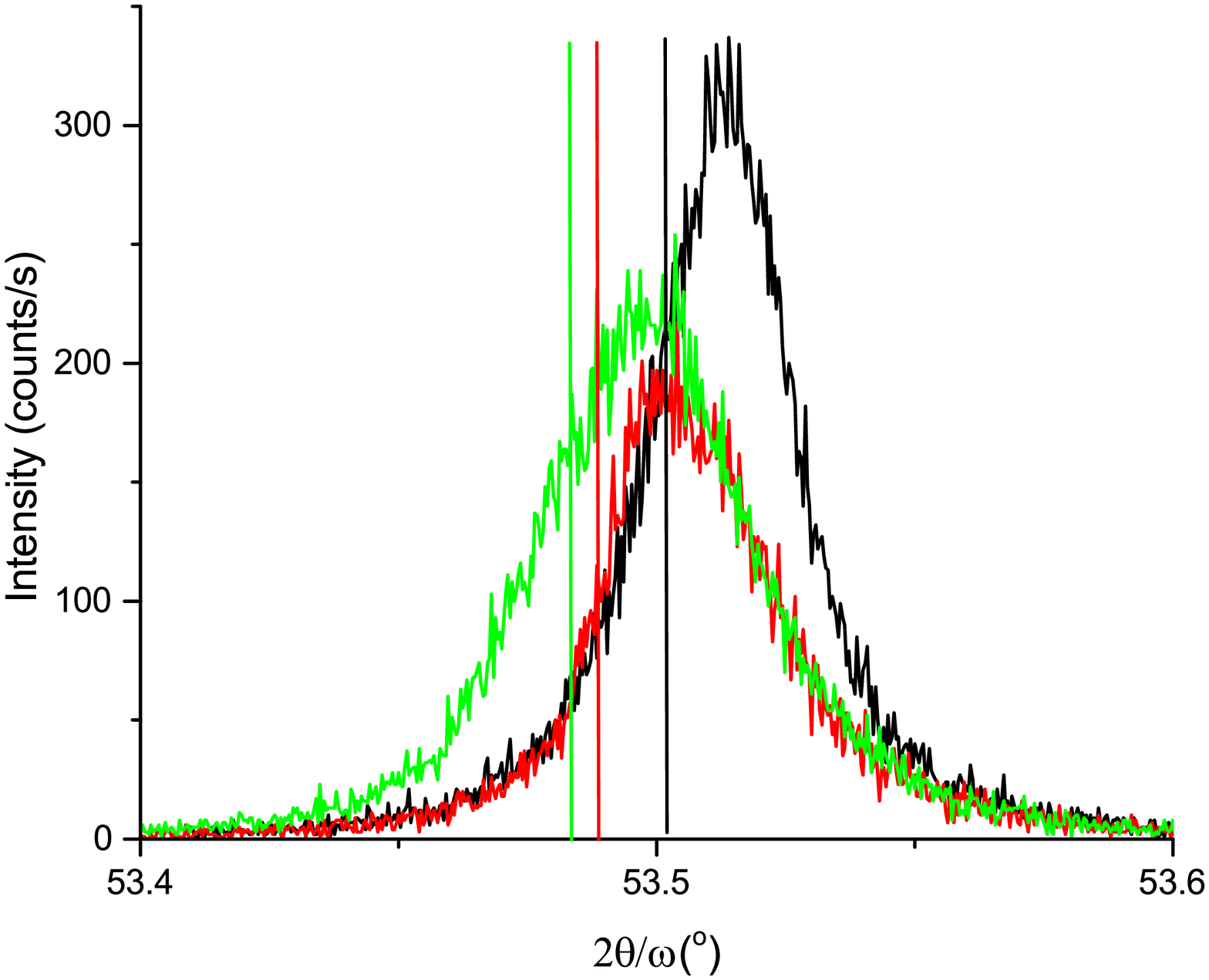}
\vspace{-0.1in}
  {\\ Fig.S6. Sample 317s. XRD (205)-reflection $2\theta/\omega$ scans.}
  \label{317sx}
\end{minipage}
\vspace{-0.1in}
\end{figure}

%\end{page}

\begin{figure}[h!]
{\bf Sample 317s7.}\\
The 317s7 ($x=0.15$) sample is strained along one of the $c$-plane crystallographic directions (discussed also in the main text). {\bf Direction of the maximal $H_{c2}$ is perpendicular to the strain axis}. Direction of maximal normal magneotresistance is rotated on 60$^\circ$ from direction of maximal $H_{c2}$. Anisotropy factor $H_{c2}^{max}/H_{c2}^{min}$ = 4.5. $\Delta a/a = 0.032\%$.\\
\vspace{-0.1in}
\begin{minipage}{.49\textwidth}
  \includegraphics[width=6.5cm]{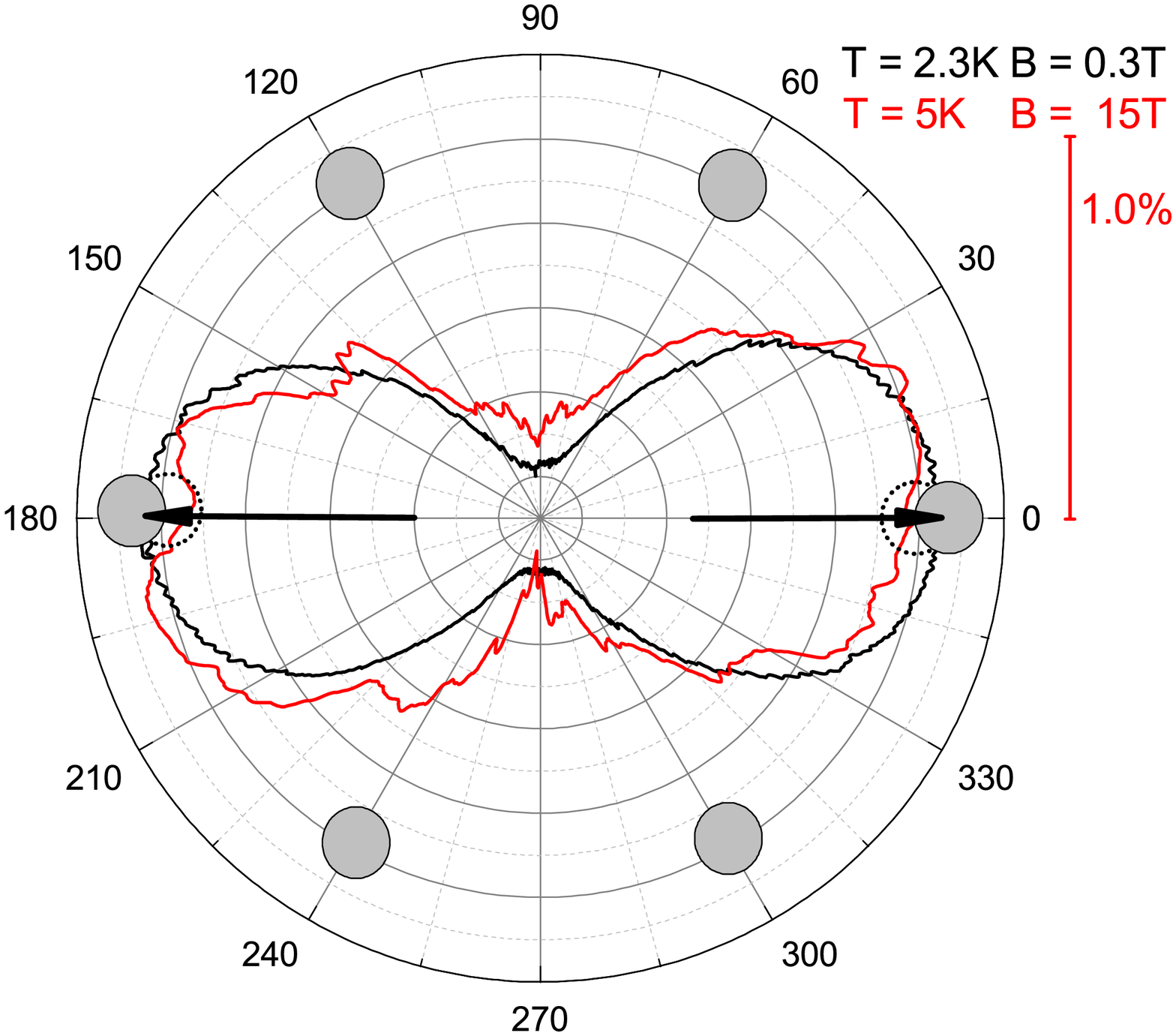}
\vspace{-0.1in}
  {\\ Fig.S7. Sample 317s7. Magnetoresistance above and below $T_c$.}
  \label{317s7}
\end{minipage}
\begin{minipage}{.49\textwidth}
  \includegraphics[width=8cm]{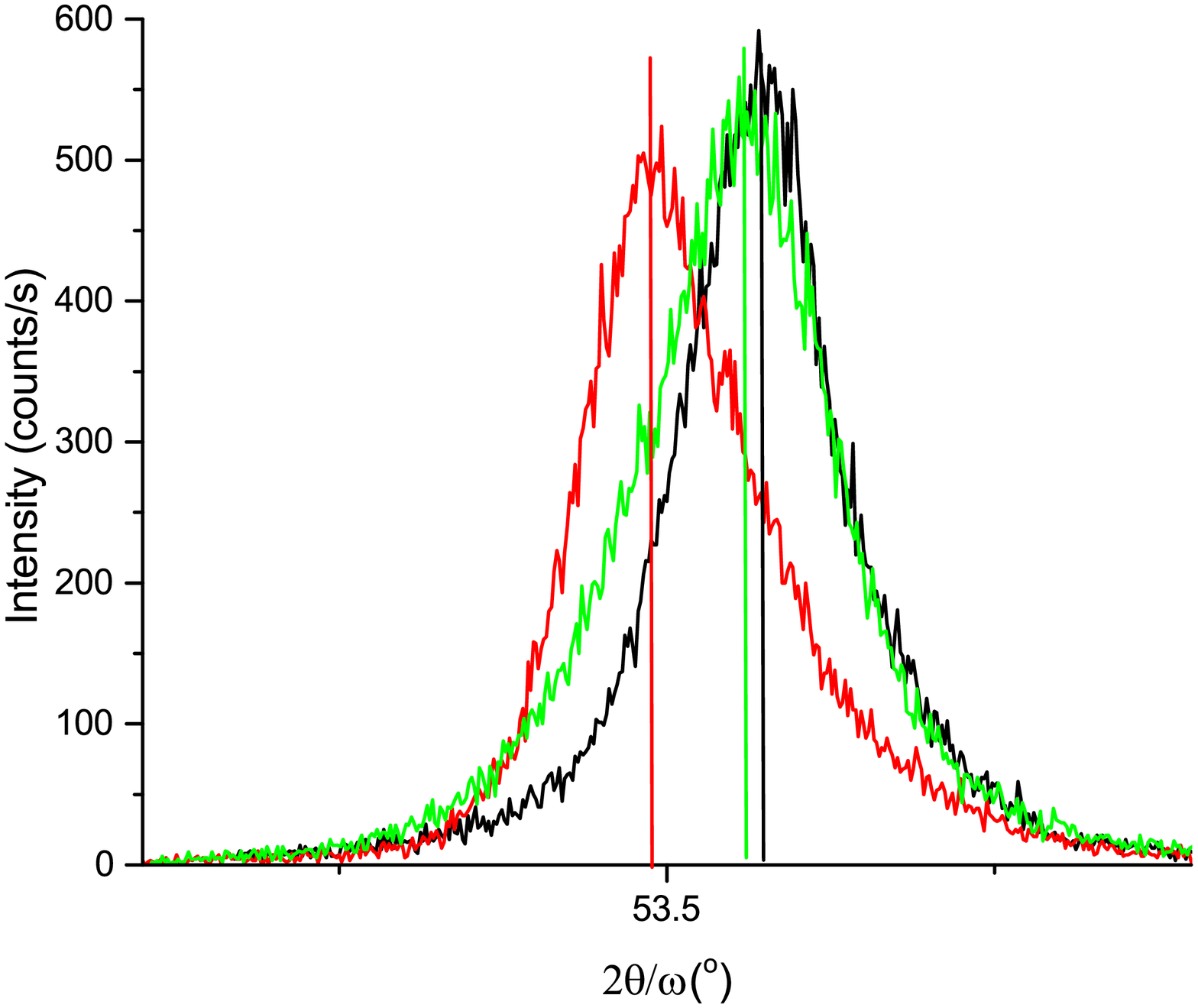}
\vspace{-0.1in}
  {\\ Fig.S8. Sample 317s7. XRD (205)-reflection $2\theta/\omega$ scans.}
  \label{317s7x}
\end{minipage}
\vspace{-0.0in}

{\bf Sample 317s9.}\\
The 317s9 ($x=0.15$) sample is not deformed within experimental precision (discussed also in the main text). {\bf Direction of the maximal $H_{c2}$ is parallel to the some crystalline axis, as if the sample was slightly compressed}. Direction of maximal normal magnetoresistance is rotated on 60$^\circ$ from direction of maximal $H_{c2}$. Anisotropy factor $H_{c2}^{max}/H_{c2}^{min}$ =1.8.

\begin{minipage}{.49\textwidth}
  \includegraphics[width=6.7cm]{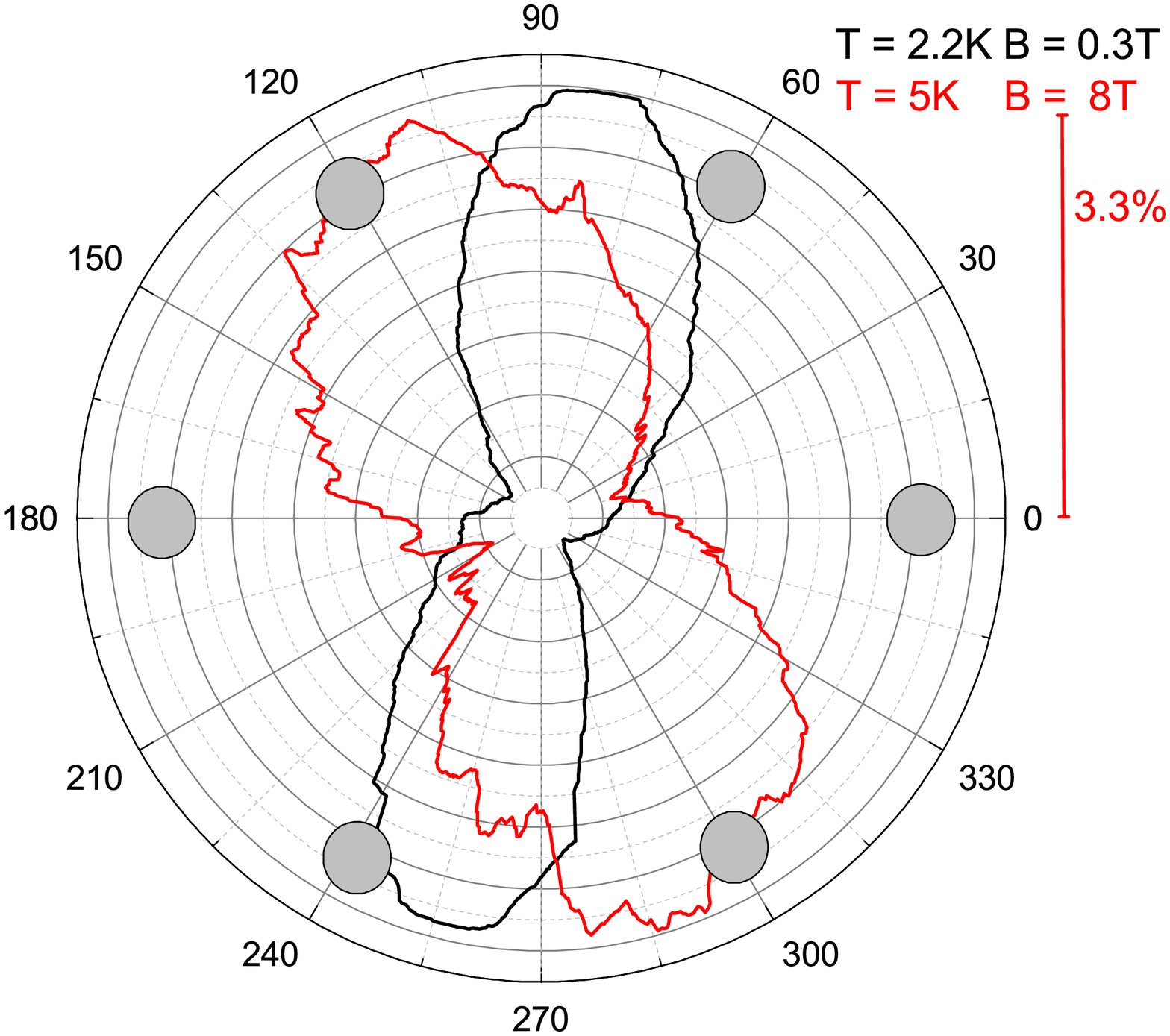}
\vspace{-0.1in}
  {\\ Fig.S9. Sample 317s9. Magnetoresistance above and below $T_c$.}
  \label{317s9}
\end{minipage}
\begin{minipage}{.49\textwidth}
  \includegraphics[width=8cm]{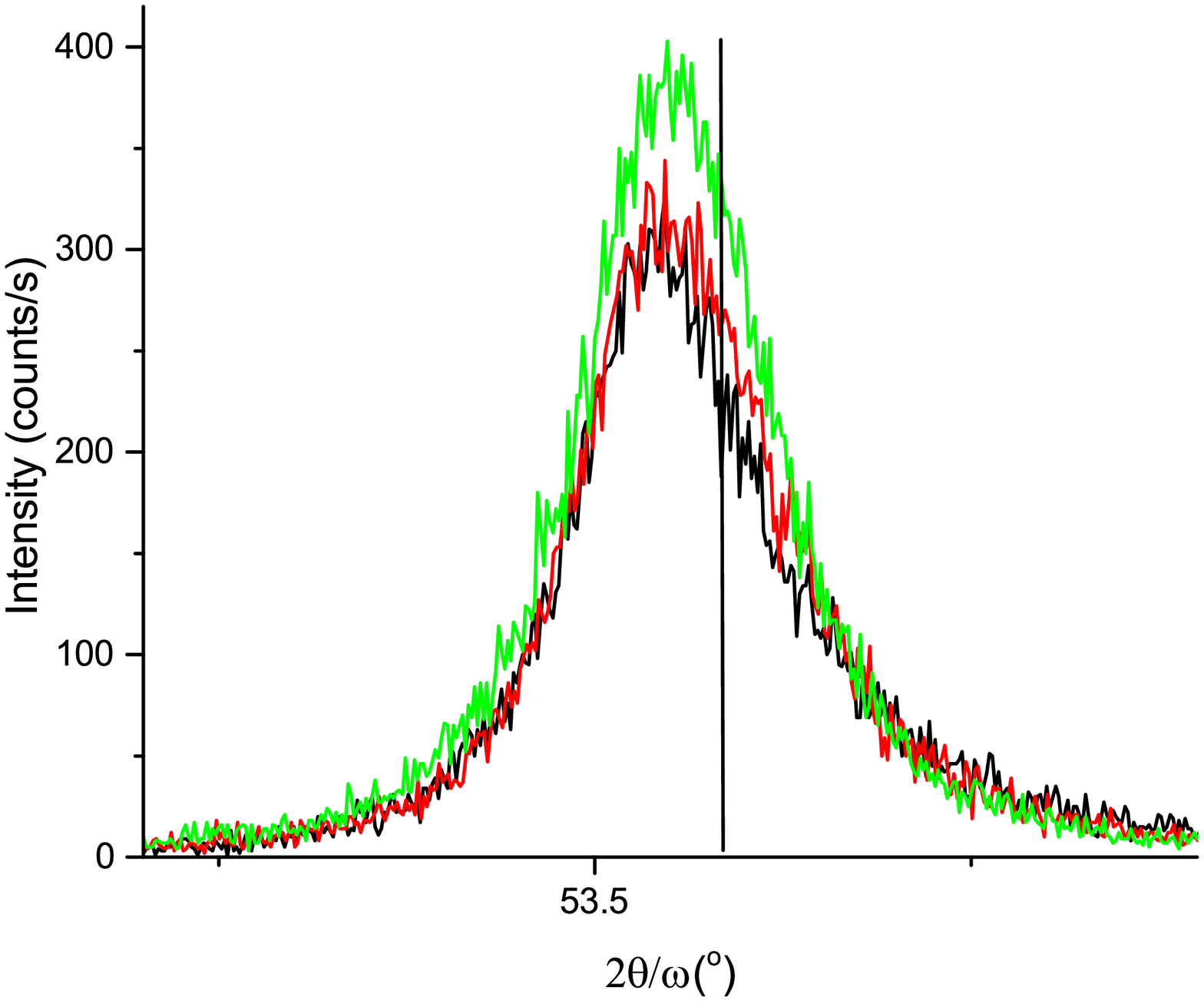}
\vspace{-0.1in}
  {\\ Fig.S10. Sample 317s9. XRD (205)-reflection $2\theta/\omega$ scans.}
  \label{317s9x}
\end{minipage}

{\bf Sample 306s4.}\\
The 306s4 ($x=0.1$) sample is compressed along one of the $c$-plane crystallographic directions. {\bf Direction of the maximal $H_{c2}$ is parallel to the compression axis}. Direction of maximal normal magnetoresistance is perpendicular to compression axis. Anisotropy factor $H_{c2}^{max}/H_{c2}^{min}\approx 3$. $\Delta a/a = 0.025\%$\\

\begin{minipage}{.49\textwidth}
  \includegraphics[width=6.4cm]{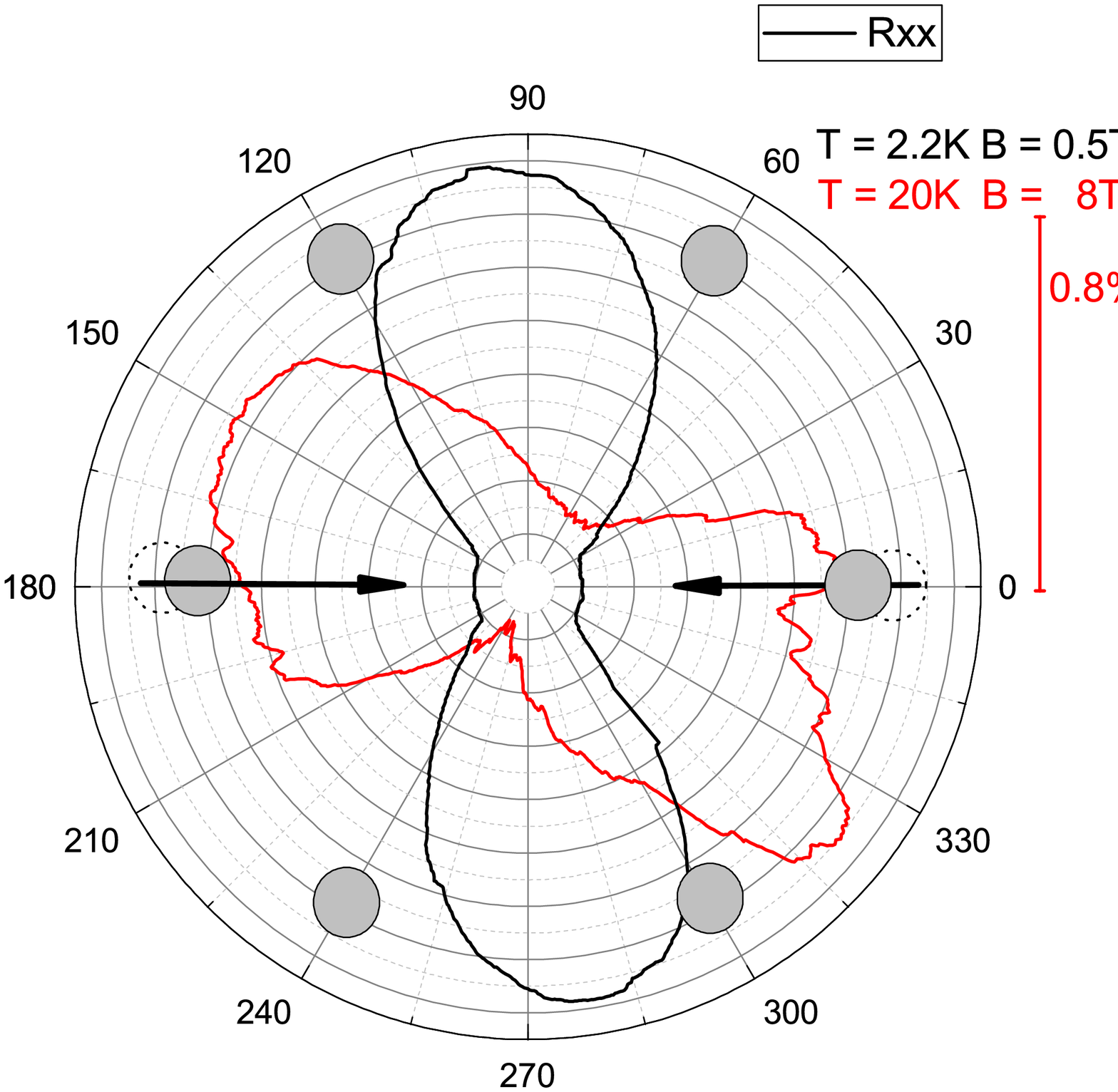}
  {\\ Fig.S11. Sample 306s4. Magnetoresistance above and below $T_c$.}
  \label{306s4}
\end{minipage}
\begin{minipage}{.49\textwidth}
  \includegraphics[width=8cm]{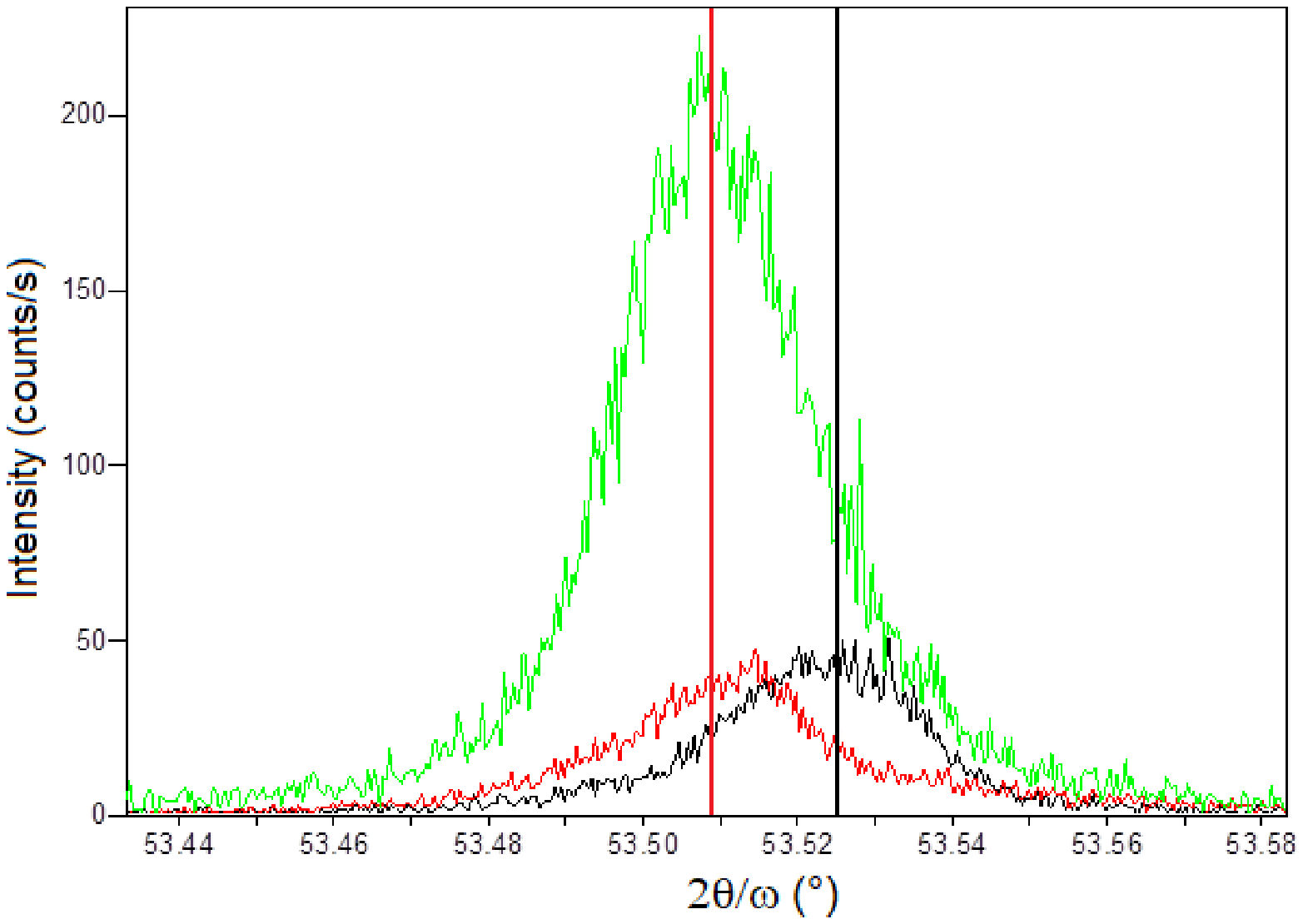}
\vspace{-0.1in}
  {\\ Fig.S12. Sample 306s4. XRD (205)-reflection $2\theta/\omega$ scans.}
  \label{306s4x}
\end{minipage}
\vspace{-0.1in}
\end{figure}

\begin{figure}[h!]
{\bf Sample 306s}.\\
The 306s ($x=0.1$) sample is compressed along one of the $c$-plane crystallographic directions. {\bf Direction of the maximal $H_{c2}$ is parallel to the compression axis}. Direction of maximal normal magnetoresistance is perpendicular to compression axis. Anisotropy factor $H_{c2}^{max}/H_{c2}^{min}$ = 3. $\Delta a/a = 0.019\%$\\

\begin{minipage}{.49\textwidth}
	\includegraphics*[trim=30 545 450 0,clip=true,width=6.5cm]{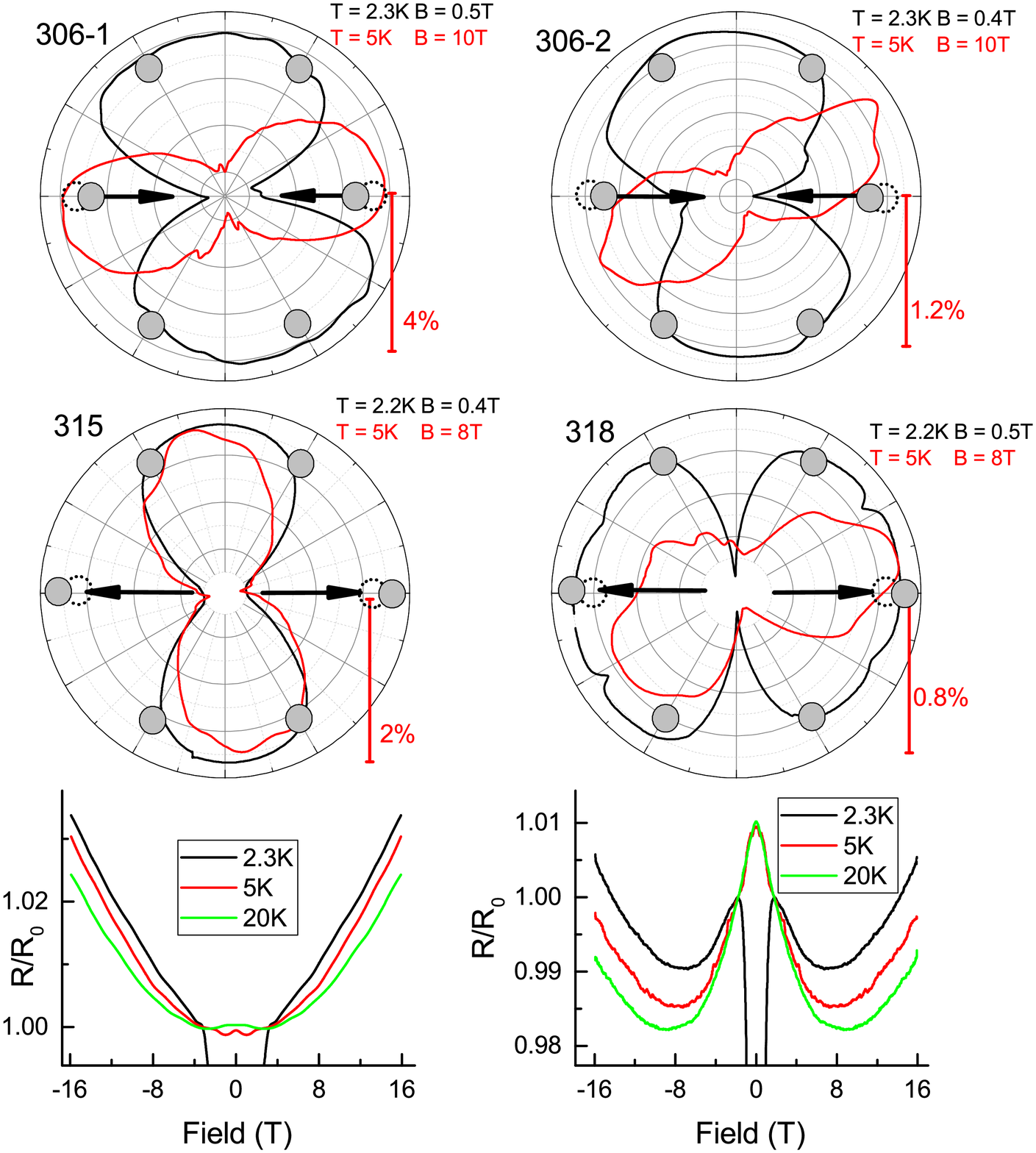}
  \vspace{-0.1in}	
	{\\ Fig.S13. Sample 306s. Magnetoresistance above and below T$_c$.}
	\label{306s}
\end{minipage}
\begin{minipage}{.49\textwidth}
  \includegraphics[width=8cm]{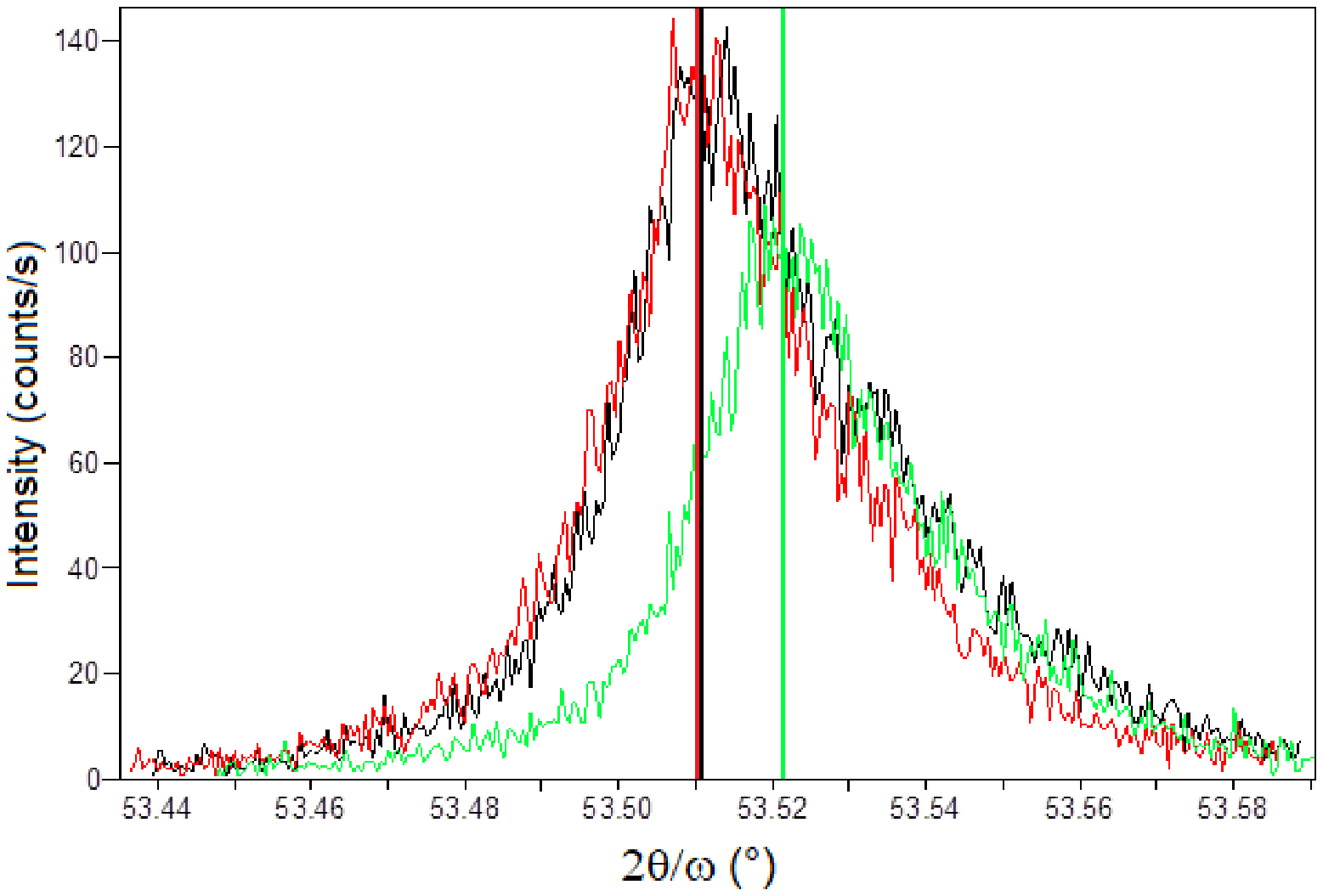}
  \vspace{-0.1in}  
  {\\ Fig.S14. Sample 306s. XRD (205)-reflection $2\theta/\omega$ scans.}
  \label{306sx}
\end{minipage}

{\bf Sample 306b.}

The 306b ($x=0.1$) sample is compressed along one of the $c$-plane crystallographic directions. {\bf Direction of the maximal $H_{c2}$ is parallel to the compression axis}. Direction of maximal normal magneotresistance is perpendicular to compression axis. Anisotropy factor $H_{c2}^{max}/H_{c2}^{min}$ = 3. $\Delta a/a = 0.019\%$\\

\begin{minipage}{.49\textwidth}
	\includegraphics*[trim=395 535 60 0,clip=true,width=6.5cm]{4samples.eps}
  \vspace{-0.1in}	
	{\\ Fig.S15. Sample 306B. Magnetoresistance above and below $T_c$.}
	\label{306B}
\end{minipage}
\begin{minipage}{.49\textwidth}
  \includegraphics[width=8cm]{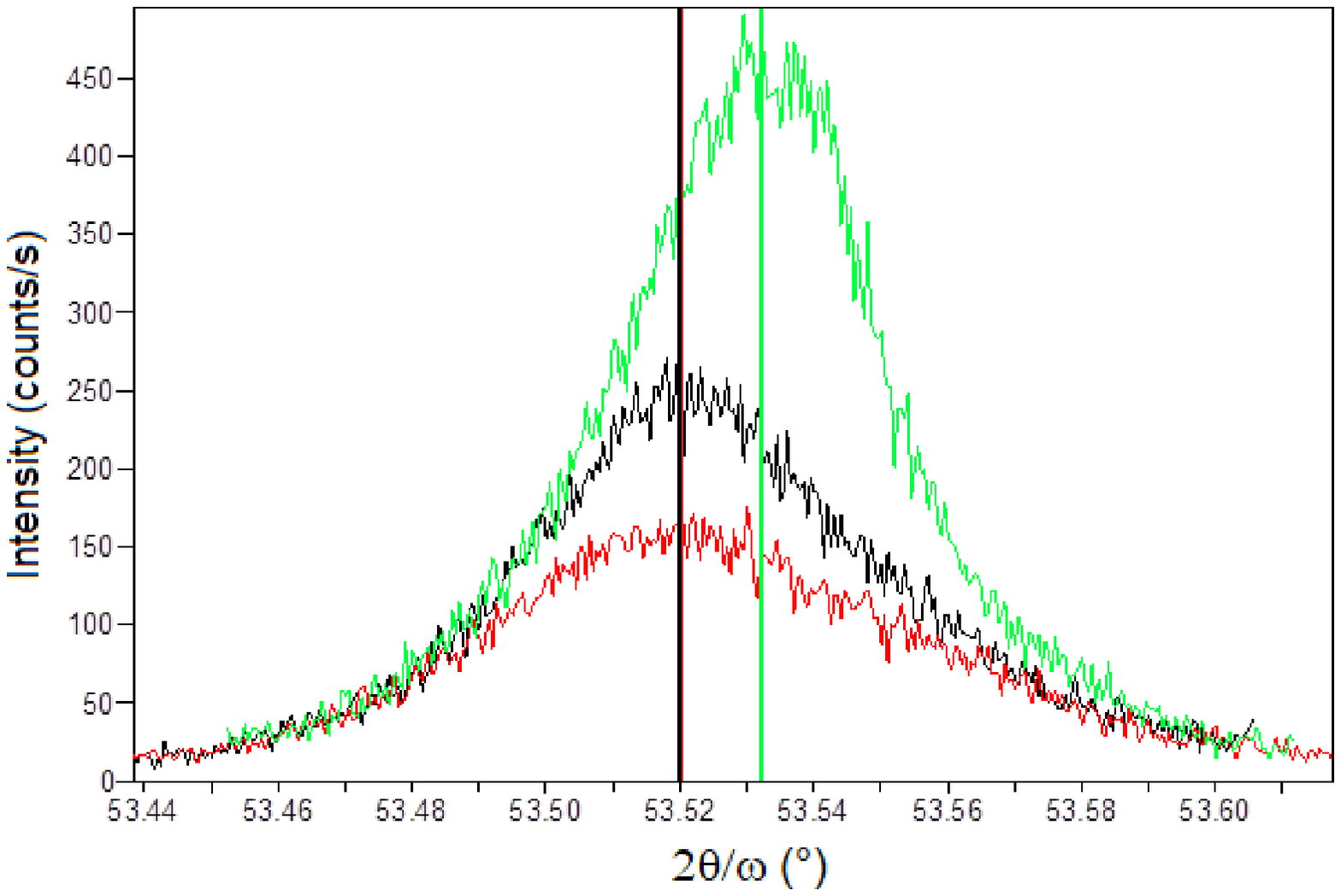}
  \vspace{-0.1in}  
  {\\ Fig.S16. Sample 306B. XRD (205)-reflection $2\theta/\omega$ scans.}
  \label{306Bx}
\end{minipage}

{\bf Sample 318.}\\
The 318 ($x=0.2$) sample is strained along one of the $c$-plane crystallographic directions. {\bf Direction of the maximal $H_{c2}$ is perpendicular to the strain axis}. Direction of maximal normal magnetoresistance is perpendicular to the direction of maximal $H_{c2}$. Anisotropy factor $H_{c2}^{max}/H_{c2}^{min}$ = 8. $\Delta a/a = 0.021\%$.

\begin{minipage}{.49\textwidth}
	\includegraphics[trim=400 255 60 280,clip=true,width=6.5cm]{4samples.eps}
  \vspace{-0.1in}
  {\\ Fig.S17. Sample 318. Magnetoresistance above and below $T_c$.}
  \label{318}
\end{minipage}
\begin{minipage}{.49\textwidth}
  \includegraphics[width=8cm]{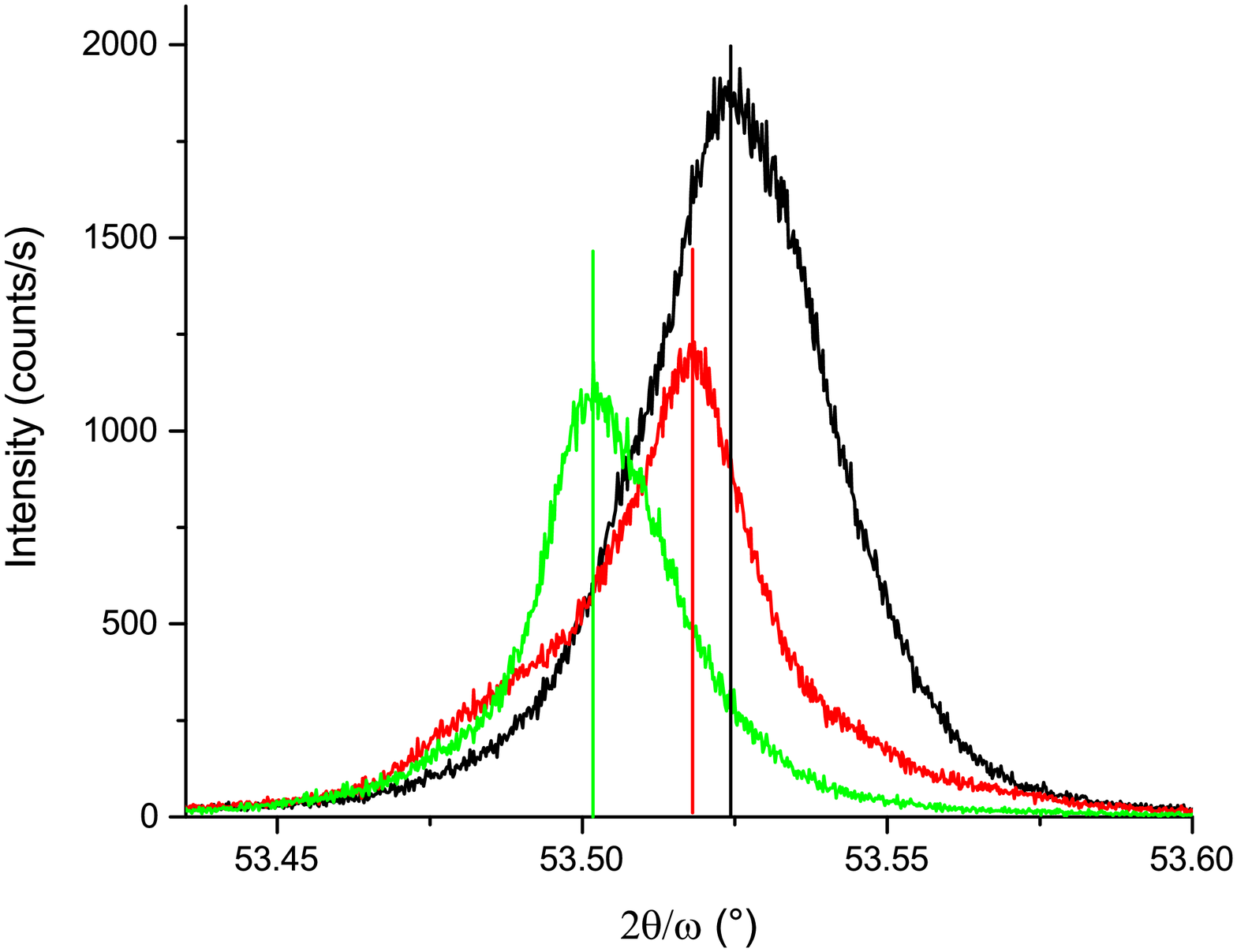}
  \vspace{-0.1in}
  {\\ Fig.S18. Sample 318. XRD (205)-reflection $2\theta/\omega$ scans.}
  \label{318x}
\end{minipage}
\end{figure}

\begin{figure}[h!]

{\bf Sample 329.}\\
The 329 sample is a co-doped Sr$_x$Cu$_y$Bi$_2$Se$_3$ sample(nominal $x=0.15$, and $y=0.01$ ). {The crystallinity of this material is not that perfect, due to effect of co-dopant (Cu ).Nevertheless, } it is strained along one of the $c$-plane crystallographic directions. {\bf Direction of the maximal $H_{c2}$ is perpendicular to the strain axis}. Direction of maximal normal magnetoresistance is perpendicular to the direction of maximal $H_{c2}$. Anisotropy factor $H_{c2}^{max}/H_{c2}^{min}\approx 3$. $\Delta a/a = 0.036\%$.\\

\begin{minipage}{.49\textwidth}
	\includegraphics[width=6.5cm]{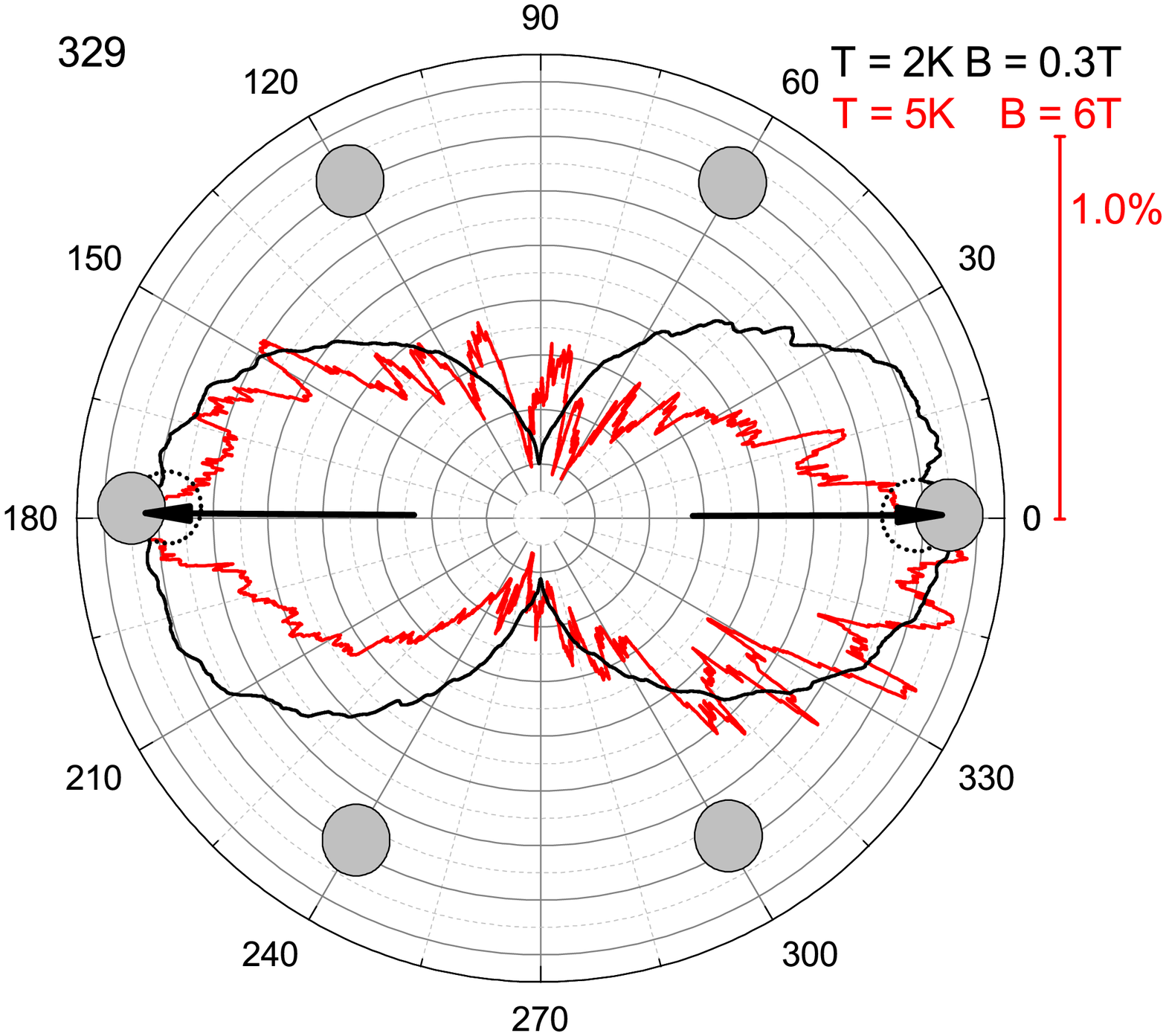}
	{\\ Fig.S19. Sample 329. Magnetoresistance above and below Tc.}
	\label{329}
\end{minipage}
\begin{minipage}{.49\textwidth}
  \includegraphics[width=8cm]{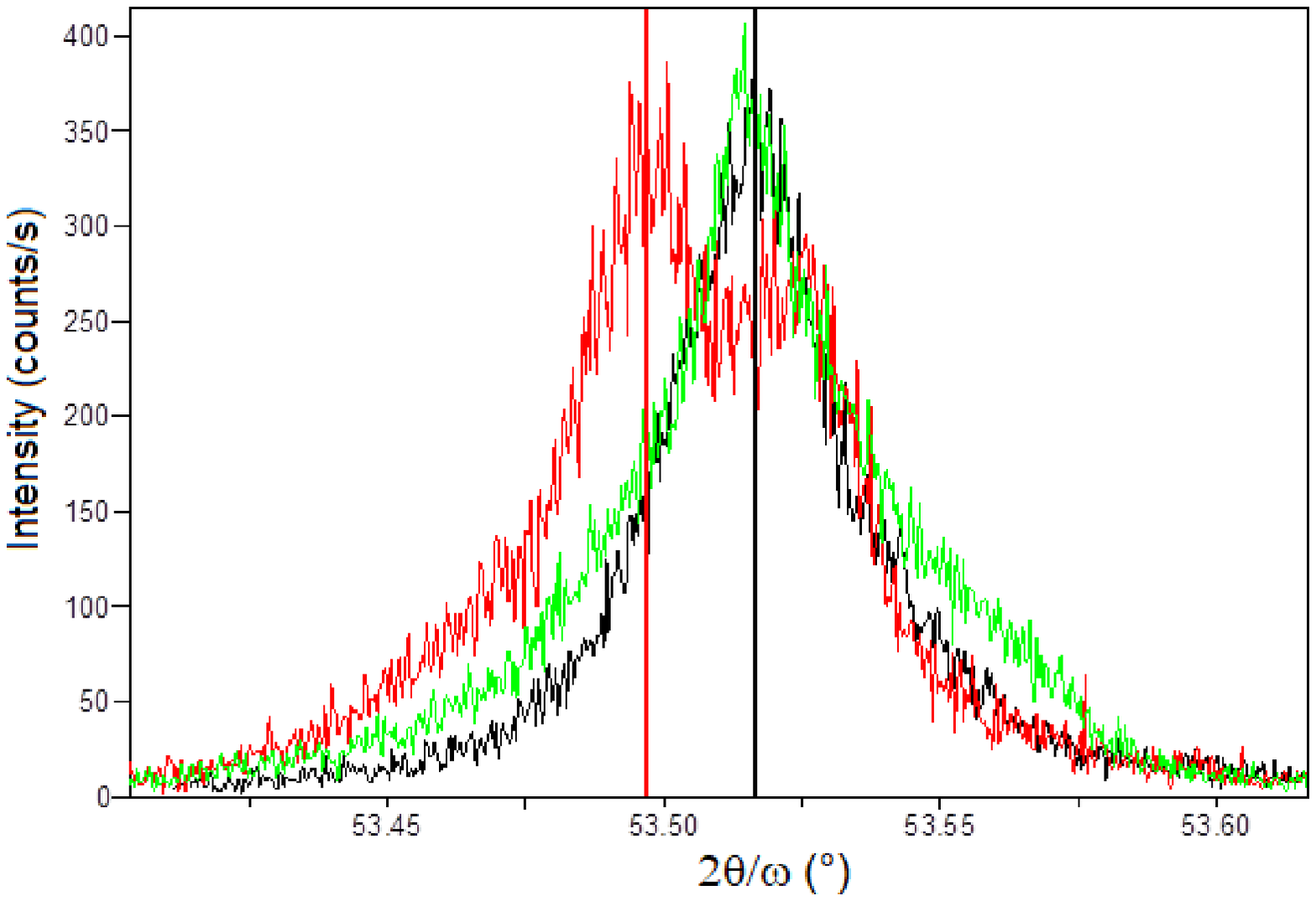}
  {\\ Fig.S20. Sample 329. XRD (205)-reflection $2\theta/\omega$ scans.}
  \label{329sx}
\end{minipage}

{\bf Sample 308}\\
The 308 ($x=0.15$)is strained mostly along one of the $c$-plane crystallographic directions. {\bf Direction of the maximal $H_{c2}$ is perpendicular to the maximal strain axis}. Normal magnetoresistance was not measured in this sample. Anisotropy factor $H_{c2}^{max}/H_{c2}^{min}$ = 6 $\Delta a/a = 0.08\%$.
Importantly, the most deformed sample demonstrates the largest $H_{c2}$ anisotropy.

\begin{minipage}{.49\textwidth}
	\includegraphics[width=6.5cm]{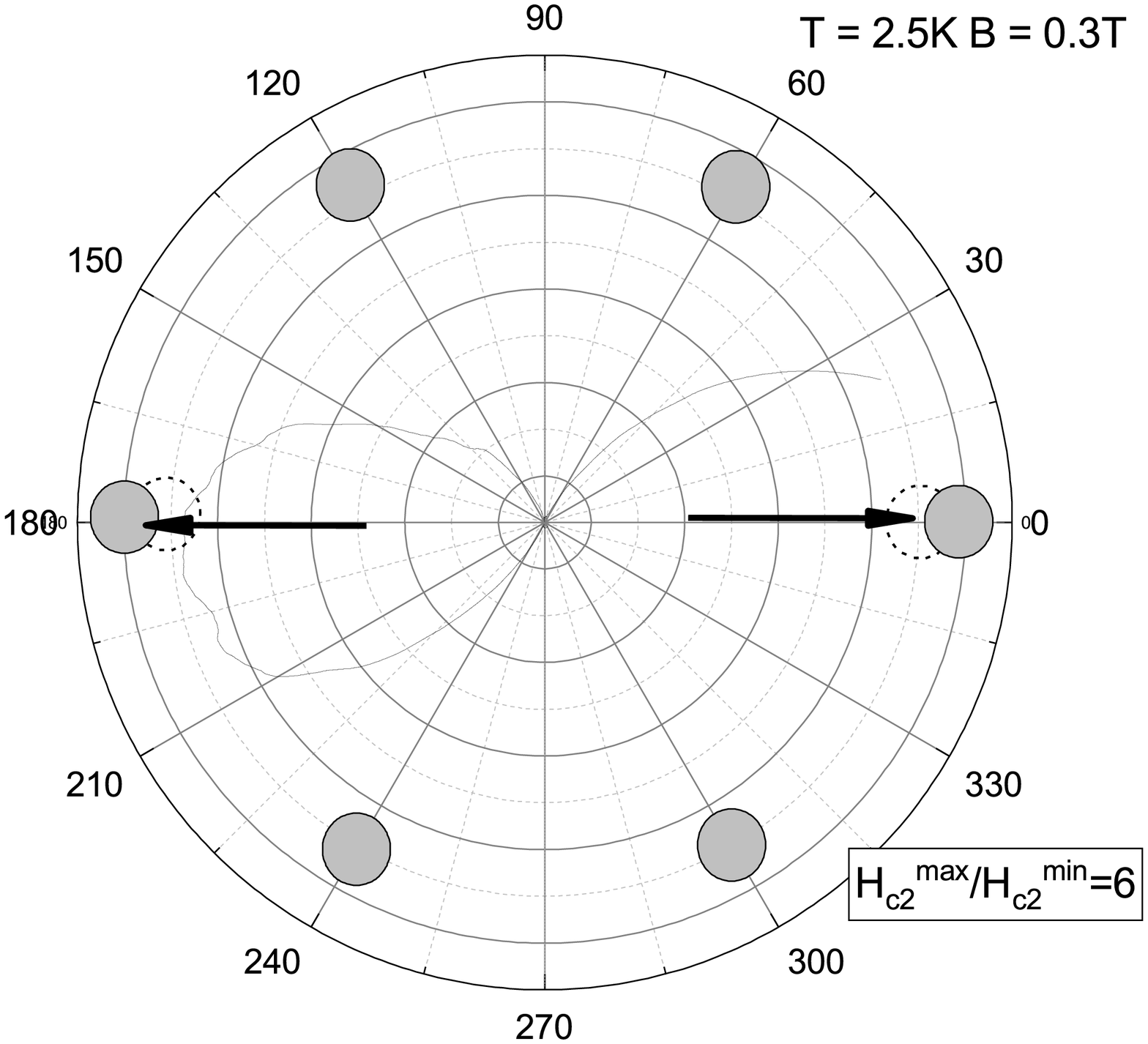}
  \vspace{-0.1in}
	{\\ Fig.S21. Sample 308. Magnetoresistance below Tc.}
	\label{308}
\end{minipage}
\begin{minipage}{.49\textwidth}
	\includegraphics[width=8cm]{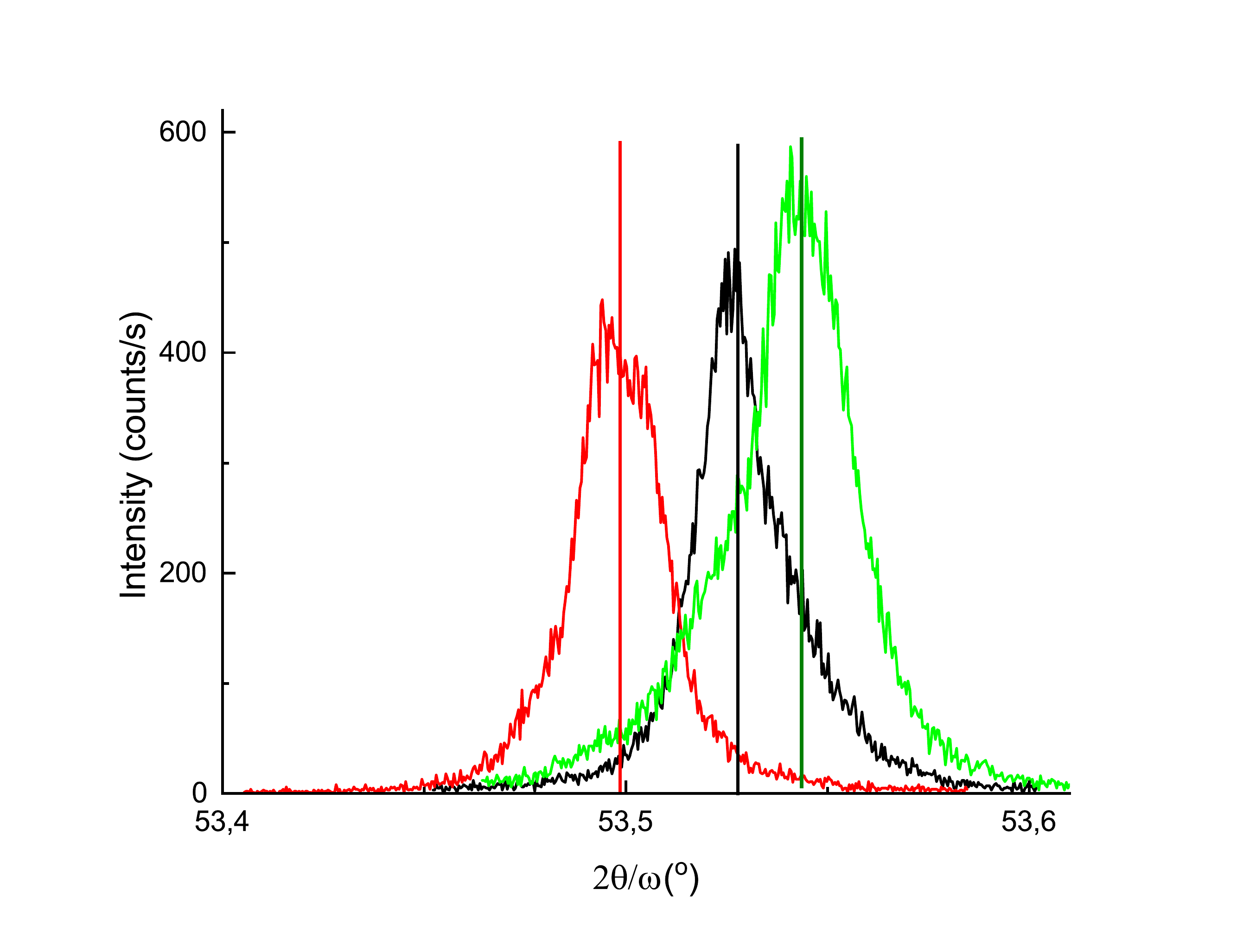}
  \vspace{-0.1in}	
	{\\ Fig.S22. Sample 308. XRD (205)-reflection $2\theta/\omega$ scans.}
	\label{308XRD}
\end{minipage}
\end{figure}

{\section{On the perfection of single crystals and absence of the twin domains.}

Multi-block structure is the the main obstacle for precise lattice parameters determination in our crystals. When the most pronounced block is under inspection, the XRD signal from the neighbouring slightly misoriented blocks might affect the 2$\theta/\omega$ curve. This may lead to mistake in reflection position determination, because the sizes and orientation of the adjacent blocks are arbitrary. 
This is the reason why XRD studies itself  require single block samples.

The boundaries between slightly misoriented blocks are nothing but arrays of dislocations. The distance between the dislocations in these arrays is determined by the angle of block misalignment.
The dislocations are in fact topological objects: they can not just disappear within the crystal and should appear at the surface. It is also extremely rare event, if the dislocations form loops or annihilate. 

Therefore the absence of rocking curve broadening or peak splitting, recorded from both sides of the thin plane-parallel crystal is a serious indication of its the single-block character and absence of grain-boundaries inside the sample.

However, our Cu $K_{\alpha}$penetration depth is approximately 10 $\mu$m, and for flat Sr$_x$Bi$_2$Se$_3$ samples (50–80 $\mu$m thickness) a possibility remains for the plane of twinning to be formed inside the crystal (lamellar twinning). Indeed, the crystal grows plane-by plane, and if such twinning plane was once formed during the growth, it will be readily expanded, as it quintuple layer step has $<90^\circ$ angle with the underlayer \cite{Sungawa}. 

In order to find the azimuthal positions of the (2~0~5) reflections and prove the absence of twins, we recorded $\phi$-scan curves on (2~0~5) reflection. As the inclination angle $\varphi_{(001)/(205)}=72.50^\circ$ exceeds the diffraction angle $\theta_{(205)}=26.85^\circ$, the $\phi$-scans were recorded in the grazing diffraction geometry, as explained in the main text.
For this measurement, the goniometer cradle with the sample should be rotated about the [0~0~1] axis, and the $\phi$-dependence of the XRD signal should be recorded.

\begin{figure}[t!]
	\center
	\includegraphics[width=13 cm]{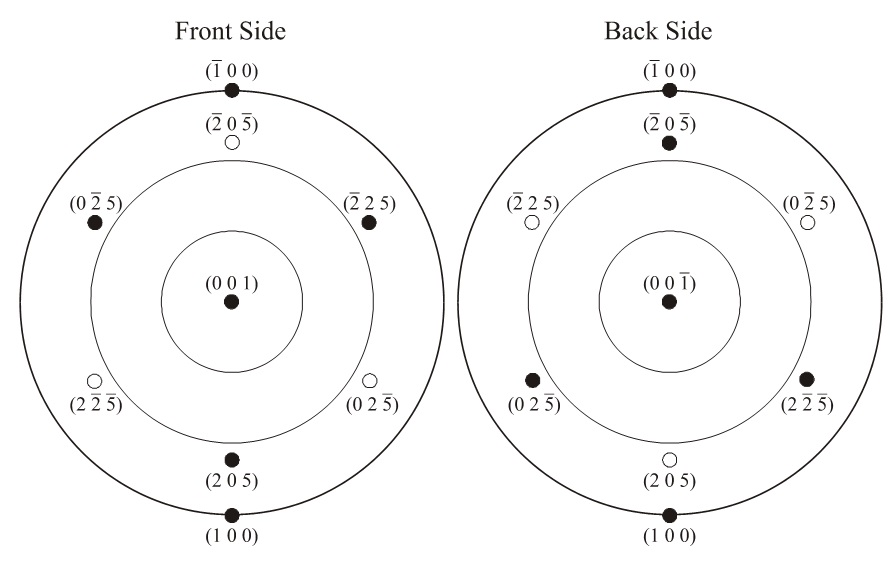}
	{\\ Fig.S25 Schematic representation of the family of (2~0~5)-reflections for $\overline{3}2/$m point symmetry group. Left panel: as observed from the top side, right panel - from the bottom side of the crystal. Black circles denote reflections directed to the top semisphere, empty circles -to the bottom semisphere.}\label{diagr}
\end{figure}
 
 Sr-doping of Bi$_2$Se$_3$ affects neither point symmetry group $\overline{3}2/$m nor spatial symmetry group R$\overline{3}$m(\#166). Fig S25, shows the stereographic projections of the normals to $\lbrace 205 \rbrace$ planes for point symmetry group $\overline{3}2/$m before and after 180$\circ$ rotation about [1~0~0] axis, respectively. Black circles correspond to normals through the top semisphere. Empty circles correspond to normals through the bottom semisphere. For a crystals with third order inversion axis, these two groups of reflections are shifted by $\Delta\phi=60^{\circ}$.  Therefore, if the sample is rotated by 180$^\circ$ rotation about [1~0~0] axis, the $\phi$- scan of (2~0~5) reflections, recorded from the back side should be shifted by 60$^\circ$. 
 
 If the twins are present within the top 10~$\mu$m of the studied samples, then the (2~0~5) reflections should appear every 60$^\circ$. Depending on the twin relationship, one triple group of reflections might be more intensive than the other one, as we observed for the epitaxial Sr:Bi$_2$Se$_3$ layers on (111)BaF$_2$substrate \cite{Volosheniuk2019}. If there exists a twinning plane parallel to the basal plane somewhere in depth of the crystal, then reflections in the crystal,  rotated by 180$^\circ$ about [100] direction, should coincide with non-rotated, contrary to the abovementioned expectation for a single-block. 
 
The top(red) curve in Figure S26 shows the $\phi$-scan for 317s7 sample.  The peaks are separated by 120$^\circ$. No peaks in between is observed (the relevant positions are shown by circles), clearly signifying the absence of twin domains in the top $\sim$10 $\mu$m. Also this curve allows to find the directions of $a$, $b$ and $i\equiv -a-b$. 

Exactly the same measurements have already been performed in Ref.\cite{Smylie2018} (Fig. 7). In our research we go further, and flip the sample via rotating it by 180$^\circ$ about [1~0~0] axis, as it is shown next to the middle (black) curve in Fig. S26.
The middle curve shows the $\phi$-scan for sample rotated that way.
Absence of peaks from twins again indicates that bottom $\sim$10 $\mu$m are also single block. A perfect $\sim$60$^\circ$ shift between the top and middle curves means that twin boundary inside the body of the crystal is also absent.
So, we detect exactly the same (with obtained precise 60$\circ$ shift) single crystal block from top and bottom measurements. Our sample is indeed single domain crystal.

The bottom(blue) curve in the same Figure shows an example of (2~0~5) reflection $\phi$-scan for large multi-block 306 sample. This curve clearly shows a fingerprint of multi-block crystal structure: variation of mutual positions of sub-peaks with $\phi$. Note, multi-block samples also do not demonstrate twin domains.

These measurements evidence for single-crystallinity and absence of twin domains in our samples.
}

\begin{figure}[t!]
	\center
	\includegraphics[width=13 cm]{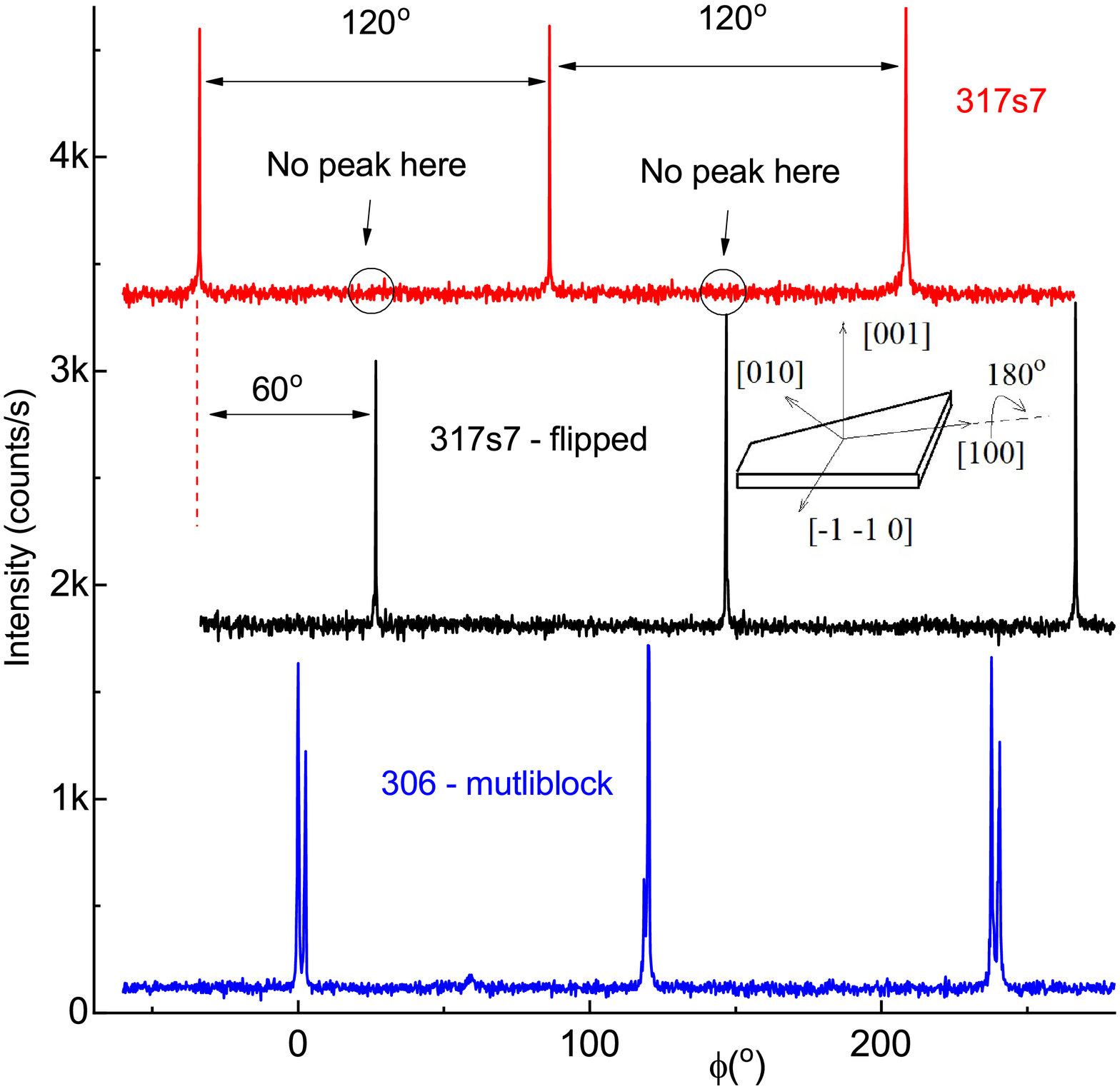}
	{\\ Fig.S26 $\phi$-scan curves. Top(red) curve - 317s7, middle(black) - the same sample rotated by 180$^\circ$ about [1~0~0] axis. Bottom curve(blue) is collected for huge multiblock sample (306 before cleavage) }\label{phiscan}
\end{figure}

\end{widetext}

\begin{thebibliography}{41}%
\makeatletter
\providecommand \@ifxundefined [1]{%
 \@ifx{#1\undefined}
}%
\providecommand \@ifnum [1]{%
 \ifnum #1\expandafter \@firstoftwo
 \else \expandafter \@secondoftwo
 \fi
}%
\providecommand \@ifx [1]{%
 \ifx #1\expandafter \@firstoftwo
 \else \expandafter \@secondoftwo
 \fi
}%
\providecommand \natexlab [1]{#1}%
\providecommand \enquote  [1]{``#1''}%
\providecommand \bibnamefont  [1]{#1}%
\providecommand \bibfnamefont [1]{#1}%
\providecommand \citenamefont [1]{#1}%
\providecommand \href@noop [0]{\@secondoftwo}%
\providecommand \href [0]{\begingroup \@sanitize@url \@href}%
\providecommand \@href[1]{\@@startlink{#1}\@@href}%
\providecommand \@@href[1]{\endgroup#1\@@endlink}%
\providecommand \@sanitize@url [0]{\catcode `\\12\catcode `\$12\catcode
  `\&12\catcode `\#12\catcode `\^12\catcode `\_12\catcode `\%12\relax}%
\providecommand \@@startlink[1]{}%
\providecommand \@@endlink[0]{}%
\providecommand \url  [0]{\begingroup\@sanitize@url \@url }%
\providecommand \@url [1]{\endgroup\@href {#1}{\urlprefix }}%
\providecommand \urlprefix  [0]{URL }%
\providecommand \Eprint [0]{\href }%
\providecommand \doibase [0]{http://dx.doi.org/}%
\providecommand \selectlanguage [0]{\@gobble}%
\providecommand \bibinfo  [0]{\@secondoftwo}%
\providecommand \bibfield  [0]{\@secondoftwo}%
\providecommand \translation [1]{[#1]}%
\providecommand \BibitemOpen [0]{}%
\providecommand \bibitemStop [0]{}%
\providecommand \bibitemNoStop [0]{.\EOS\space}%
\providecommand \EOS [0]{\spacefactor3000\relax}%
\providecommand \BibitemShut  [1]{\csname bibitem#1\endcsname}%
\let\auto@bib@innerbib\@empty
%</preamble>
\bibitem [{\citenamefont {Fu}\ and\ \citenamefont {Berg}(2010)}]{Fu2010}%
  \BibitemOpen
  \bibfield  {author} {\bibinfo {author} {\bibfnamefont {L.}~\bibnamefont
  {Fu}}\ and\ \bibinfo {author} {\bibfnamefont {E.}~\bibnamefont {Berg}},\
  }\href {https://link.aps.org/doi/10.1103/PhysRevLett.105.097001} {\bibfield
  {journal} {\bibinfo  {journal} {Phys. Rev. Lett.}\ }\textbf {\bibinfo
  {volume} {105}},\ \bibinfo {pages} {097001} (\bibinfo {year}
  {2010})}\BibitemShut {NoStop}%
\bibitem [{\citenamefont {Fu}\ and\ \citenamefont {Kane}(2008)}]{Fu2008}%
  \BibitemOpen
  \bibfield  {author} {\bibinfo {author} {\bibfnamefont {L.}~\bibnamefont
  {Fu}}\ and\ \bibinfo {author} {\bibfnamefont {C.~L.}\ \bibnamefont {Kane}},\
  }\href {https://link.aps.org/doi/10.1103/PhysRevLett.100.096407} {\bibfield
  {journal} {\bibinfo  {journal} {Phys. Rev. Lett.}\ }\textbf {\bibinfo
  {volume} {100}},\ \bibinfo {pages} {096407} (\bibinfo {year}
  {2008})}\BibitemShut {NoStop}%
\bibitem [{\citenamefont {Potter}\ and\ \citenamefont {Fu}(2013)}]{Potter2013}%
  \BibitemOpen
  \bibfield  {author} {\bibinfo {author} {\bibfnamefont {A.~C.}\ \bibnamefont
  {Potter}}\ and\ \bibinfo {author} {\bibfnamefont {L.}~\bibnamefont {Fu}},\
  }\href {https://link.aps.org/doi/10.1103/PhysRevB.88.121109} {\bibfield
  {journal} {\bibinfo  {journal} {Phys. Rev. B}\ }\textbf {\bibinfo {volume}
  {88}},\ \bibinfo {pages} {121109(R)} (\bibinfo {year} {2013})}\BibitemShut
  {NoStop}%
\bibitem [{\citenamefont {Sarma}\ \emph {et~al.}(2015)\citenamefont {Sarma},
  \citenamefont {Freedman},\ and\ \citenamefont {Nayak}}]{Sarma2015}%
  \BibitemOpen
  \bibfield  {author} {\bibinfo {author} {\bibfnamefont {S.~D.}\ \bibnamefont
  {Sarma}}, \bibinfo {author} {\bibfnamefont {M.}~\bibnamefont {Freedman}}, \
  and\ \bibinfo {author} {\bibfnamefont {C.}~\bibnamefont {Nayak}},\ }\href
  {https://doi.org/10.1038/npjqi.2015.1} {\bibfield  {journal} {\bibinfo
  {journal} {Npj Quantum Information}\ }\textbf {\bibinfo {volume} {1}},\
  \bibinfo {pages} {15001} (\bibinfo {year} {2015})}\BibitemShut {NoStop}%
\bibitem [{\citenamefont {Elliott}\ and\ \citenamefont
  {Franz}(2015)}]{Elliott2015}%
  \BibitemOpen
  \bibfield  {author} {\bibinfo {author} {\bibfnamefont {S.~R.}\ \bibnamefont
  {Elliott}}\ and\ \bibinfo {author} {\bibfnamefont {M.}~\bibnamefont
  {Franz}},\ }\href {https://link.aps.org/doi/10.1103/RevModPhys.87.137}
  {\bibfield  {journal} {\bibinfo  {journal} {Rev. Mod. Phys.}\ }\textbf
  {\bibinfo {volume} {87}},\ \bibinfo {pages} {137 } (\bibinfo {year}
  {2015})}\BibitemShut {NoStop}%
\bibitem [{\citenamefont {Akzyanov}\ \emph {et~al.}(2015)\citenamefont
  {Akzyanov}, \citenamefont {Rakhmanov}, \citenamefont {Rozhkov},\ and\
  \citenamefont {Nori}}]{Akzyanov2015}%
  \BibitemOpen
  \bibfield  {author} {\bibinfo {author} {\bibfnamefont {R.~S.}\ \bibnamefont
  {Akzyanov}}, \bibinfo {author} {\bibfnamefont {A.~L.}\ \bibnamefont
  {Rakhmanov}}, \bibinfo {author} {\bibfnamefont {A.~V.}\ \bibnamefont
  {Rozhkov}}, \ and\ \bibinfo {author} {\bibfnamefont {F.}~\bibnamefont
  {Nori}},\ }\href {https://link.aps.org/doi/10.1103/PhysRevB.92.075432}
  {\bibfield  {journal} {\bibinfo  {journal} {Phys. Rev. B}\ }\textbf {\bibinfo
  {volume} {92}},\ \bibinfo {pages} {075432} (\bibinfo {year}
  {2015})}\BibitemShut {NoStop}%
\bibitem [{\citenamefont {Akzyanov}\ \emph {et~al.}(2016)\citenamefont
  {Akzyanov}, \citenamefont {Rakhmanov}, \citenamefont {Rozhkov},\ and\
  \citenamefont {Nori}}]{Akzyanov2016}%
  \BibitemOpen
  \bibfield  {author} {\bibinfo {author} {\bibfnamefont {R.~S.}\ \bibnamefont
  {Akzyanov}}, \bibinfo {author} {\bibfnamefont {A.~L.}\ \bibnamefont
  {Rakhmanov}}, \bibinfo {author} {\bibfnamefont {A.~V.}\ \bibnamefont
  {Rozhkov}}, \ and\ \bibinfo {author} {\bibfnamefont {F.}~\bibnamefont
  {Nori}},\ }\href {https://link.aps.org/doi/10.1103/PhysRevB.94.125428}
  {\bibfield  {journal} {\bibinfo  {journal} {Phys. Rev. B}\ }\textbf {\bibinfo
  {volume} {94}},\ \bibinfo {pages} {125428} (\bibinfo {year}
  {2016})}\BibitemShut {NoStop}%
\bibitem [{\citenamefont {Sato}\ and\ \citenamefont {Ando}(2017)}]{Sato2017}%
  \BibitemOpen
  \bibfield  {author} {\bibinfo {author} {\bibfnamefont {M.}~\bibnamefont
  {Sato}}\ and\ \bibinfo {author} {\bibfnamefont {Y.}~\bibnamefont {Ando}},\
  }\href {http://dx.doi.org/10.1088/1361-6633/aa6ac7} {\bibfield  {journal}
  {\bibinfo  {journal} {Reports on Progress in Physics}\ }\textbf {\bibinfo
  {volume} {80}},\ \bibinfo {pages} {076501} (\bibinfo {year}
  {2017})}\BibitemShut {NoStop}%
\bibitem [{\citenamefont {Zhang}\ \emph {et~al.}(2009)\citenamefont {Zhang},
  \citenamefont {Liu}, \citenamefont {Qi}, \citenamefont {Dai}, \citenamefont
  {Fang},\ and\ \citenamefont {Zhang}}]{Zhang2009}%
  \BibitemOpen
  \bibfield  {author} {\bibinfo {author} {\bibfnamefont {H.}~\bibnamefont
  {Zhang}}, \bibinfo {author} {\bibfnamefont {C.-X.}\ \bibnamefont {Liu}},
  \bibinfo {author} {\bibfnamefont {X.-L.}\ \bibnamefont {Qi}}, \bibinfo
  {author} {\bibfnamefont {X.}~\bibnamefont {Dai}}, \bibinfo {author}
  {\bibfnamefont {Z.}~\bibnamefont {Fang}}, \ and\ \bibinfo {author}
  {\bibfnamefont {S.-C.}\ \bibnamefont {Zhang}},\ }\href
  {https://doi.org/10.1038/nphys1270} {\bibfield  {journal} {\bibinfo
  {journal} {Nature Physics}\ }\textbf {\bibinfo {volume} {5}},\ \bibinfo
  {pages} {438} (\bibinfo {year} {2009})}\BibitemShut {NoStop}%
\bibitem [{\citenamefont {Hasan}\ and\ \citenamefont {Kane}(2010)}]{Hasan2010}%
  \BibitemOpen
  \bibfield  {author} {\bibinfo {author} {\bibfnamefont {M.~Z.}\ \bibnamefont
  {Hasan}}\ and\ \bibinfo {author} {\bibfnamefont {C.~L.}\ \bibnamefont
  {Kane}},\ }\href {https://link.aps.org/doi/10.1103/RevModPhys.82.3045}
  {\bibfield  {journal} {\bibinfo  {journal} {Rev. Mod. Phys.}\ }\textbf
  {\bibinfo {volume} {82}},\ \bibinfo {pages} {3045} (\bibinfo {year}
  {2010})}\BibitemShut {NoStop}%
\bibitem [{\citenamefont {Sau}\ \emph {et~al.}(2010)\citenamefont {Sau},
  \citenamefont {Lutchyn}, \citenamefont {Tewari},\ and\ \citenamefont
  {Das~Sarma}}]{Sau2010}%
  \BibitemOpen
  \bibfield  {author} {\bibinfo {author} {\bibfnamefont {J.~D.}\ \bibnamefont
  {Sau}}, \bibinfo {author} {\bibfnamefont {R.~M.}\ \bibnamefont {Lutchyn}},
  \bibinfo {author} {\bibfnamefont {S.}~\bibnamefont {Tewari}}, \ and\ \bibinfo
  {author} {\bibfnamefont {S.}~\bibnamefont {Das~Sarma}},\ }\href {\doibase
  10.1103/PhysRevB.82.094522} {\bibfield  {journal} {\bibinfo  {journal} {Phys.
  Rev. B}\ }\textbf {\bibinfo {volume} {82}},\ \bibinfo {pages} {094522}
  (\bibinfo {year} {2010})}\BibitemShut {NoStop}%
\bibitem [{\citenamefont {Wray}\ \emph {et~al.}(2010)\citenamefont {Wray},
  \citenamefont {Xu}, \citenamefont {Xia}, \citenamefont {Hor}, \citenamefont
  {Qian}, \citenamefont {Fedorov}, \citenamefont {Lin}, \citenamefont {Bansil},
  \citenamefont {Cava},\ and\ \citenamefont {Hasan}}]{Wray2010}%
  \BibitemOpen
  \bibfield  {author} {\bibinfo {author} {\bibfnamefont {L.~A.}\ \bibnamefont
  {Wray}}, \bibinfo {author} {\bibfnamefont {S.-Y.}\ \bibnamefont {Xu}},
  \bibinfo {author} {\bibfnamefont {Y.}~\bibnamefont {Xia}}, \bibinfo {author}
  {\bibfnamefont {Y.~S.}\ \bibnamefont {Hor}}, \bibinfo {author} {\bibfnamefont
  {D.}~\bibnamefont {Qian}}, \bibinfo {author} {\bibfnamefont {A.~V.}\
  \bibnamefont {Fedorov}}, \bibinfo {author} {\bibfnamefont {H.}~\bibnamefont
  {Lin}}, \bibinfo {author} {\bibfnamefont {A.}~\bibnamefont {Bansil}},
  \bibinfo {author} {\bibfnamefont {R.~J.}\ \bibnamefont {Cava}}, \ and\
  \bibinfo {author} {\bibfnamefont {M.~Z.}\ \bibnamefont {Hasan}},\ }\href
  {https://doi.org/10.1038/nphys1762} {\bibfield  {journal} {\bibinfo
  {journal} {Nature Physics}\ }\textbf {\bibinfo {volume} {6}},\ \bibinfo
  {pages} {855} (\bibinfo {year} {2010})}\BibitemShut {NoStop}%
\bibitem [{\citenamefont {Hor}\ \emph {et~al.}(2010)\citenamefont {Hor},
  \citenamefont {Williams}, \citenamefont {Checkelsky}, \citenamefont
  {Roushan}, \citenamefont {Seo}, \citenamefont {Xu}, \citenamefont
  {Zandbergen}, \citenamefont {Yazdani}, \citenamefont {Ong},\ and\
  \citenamefont {Cava}}]{Hor2010}%
  \BibitemOpen
  \bibfield  {author} {\bibinfo {author} {\bibfnamefont {Y.~S.}\ \bibnamefont
  {Hor}}, \bibinfo {author} {\bibfnamefont {A.~J.}\ \bibnamefont {Williams}},
  \bibinfo {author} {\bibfnamefont {J.~G.}\ \bibnamefont {Checkelsky}},
  \bibinfo {author} {\bibfnamefont {P.}~\bibnamefont {Roushan}}, \bibinfo
  {author} {\bibfnamefont {J.}~\bibnamefont {Seo}}, \bibinfo {author}
  {\bibfnamefont {Q.}~\bibnamefont {Xu}}, \bibinfo {author} {\bibfnamefont
  {H.~W.}\ \bibnamefont {Zandbergen}}, \bibinfo {author} {\bibfnamefont
  {A.}~\bibnamefont {Yazdani}}, \bibinfo {author} {\bibfnamefont {N.~P.}\
  \bibnamefont {Ong}}, \ and\ \bibinfo {author} {\bibfnamefont {R.~J.}\
  \bibnamefont {Cava}},\ }\href
  {https://link.aps.org/doi/10.1103/PhysRevLett.104.057001} {\bibfield
  {journal} {\bibinfo  {journal} {Phys. Rev. Lett.}\ }\textbf {\bibinfo
  {volume} {104}},\ \bibinfo {pages} {057001} (\bibinfo {year}
  {2010})}\BibitemShut {NoStop}%
\bibitem [{\citenamefont {Sasaki}\ \emph {et~al.}(2011)\citenamefont {Sasaki},
  \citenamefont {Kriener}, \citenamefont {Segawa}, \citenamefont {Yada},
  \citenamefont {Tanaka}, \citenamefont {Sato},\ and\ \citenamefont
  {Ando}}]{Sasaki2011}%
  \BibitemOpen
  \bibfield  {author} {\bibinfo {author} {\bibfnamefont {S.}~\bibnamefont
  {Sasaki}}, \bibinfo {author} {\bibfnamefont {M.}~\bibnamefont {Kriener}},
  \bibinfo {author} {\bibfnamefont {K.}~\bibnamefont {Segawa}}, \bibinfo
  {author} {\bibfnamefont {K.}~\bibnamefont {Yada}}, \bibinfo {author}
  {\bibfnamefont {Y.}~\bibnamefont {Tanaka}}, \bibinfo {author} {\bibfnamefont
  {M.}~\bibnamefont {Sato}}, \ and\ \bibinfo {author} {\bibfnamefont
  {Y.}~\bibnamefont {Ando}},\ }\href
  {https://link.aps.org/doi/10.1103/PhysRevLett.107.217001} {\bibfield
  {journal} {\bibinfo  {journal} {Phys. Rev. Lett.}\ }\textbf {\bibinfo
  {volume} {107}},\ \bibinfo {pages} {217001} (\bibinfo {year}
  {2011})}\BibitemShut {NoStop}%
\bibitem [{\citenamefont {Matano}\ \emph {et~al.}(2016)\citenamefont {Matano},
  \citenamefont {Kriener}, \citenamefont {Segawa}, \citenamefont {Ando},\ and\
  \citenamefont {Zheng}}]{Matano2016}%
  \BibitemOpen
  \bibfield  {author} {\bibinfo {author} {\bibfnamefont {K.}~\bibnamefont
  {Matano}}, \bibinfo {author} {\bibfnamefont {M.}~\bibnamefont {Kriener}},
  \bibinfo {author} {\bibfnamefont {K.}~\bibnamefont {Segawa}}, \bibinfo
  {author} {\bibfnamefont {Y.}~\bibnamefont {Ando}}, \ and\ \bibinfo {author}
  {\bibfnamefont {G.-q.}\ \bibnamefont {Zheng}},\ }\href
  {https://doi.org/10.1038/nphys3781} {\bibfield  {journal} {\bibinfo
  {journal} {Nature Physics}\ }\textbf {\bibinfo {volume} {12}},\ \bibinfo
  {pages} {852} (\bibinfo {year} {2016})}\BibitemShut {NoStop}%
\bibitem [{\citenamefont {Shen}\ \emph {et~al.}(2017)\citenamefont {Shen},
  \citenamefont {He}, \citenamefont {Yuan}, \citenamefont {Huang},
  \citenamefont {Cho}, \citenamefont {Lee}, \citenamefont {Hor}, \citenamefont
  {Law},\ and\ \citenamefont {Lortz}}]{Shen2017}%
  \BibitemOpen
  \bibfield  {author} {\bibinfo {author} {\bibfnamefont {J.}~\bibnamefont
  {Shen}}, \bibinfo {author} {\bibfnamefont {W.-Y.}\ \bibnamefont {He}},
  \bibinfo {author} {\bibfnamefont {N.~F.~Q.}\ \bibnamefont {Yuan}}, \bibinfo
  {author} {\bibfnamefont {Z.}~\bibnamefont {Huang}}, \bibinfo {author}
  {\bibfnamefont {C.-w.}\ \bibnamefont {Cho}}, \bibinfo {author} {\bibfnamefont
  {S.~H.}\ \bibnamefont {Lee}}, \bibinfo {author} {\bibfnamefont {Y.~S.}\
  \bibnamefont {Hor}}, \bibinfo {author} {\bibfnamefont {K.~T.}\ \bibnamefont
  {Law}}, \ and\ \bibinfo {author} {\bibfnamefont {R.}~\bibnamefont {Lortz}},\
  }\href {https://doi.org/10.1038/s41535-017-0064-1} {\bibfield  {journal}
  {\bibinfo  {journal} {npj Quantum Materials}\ }\textbf {\bibinfo {volume}
  {2}},\ \bibinfo {pages} {59} (\bibinfo {year} {2017})}\BibitemShut {NoStop}%
\bibitem [{\citenamefont {Du}\ \emph {et~al.}(2017)\citenamefont {Du},
  \citenamefont {Li}, \citenamefont {Schneeloch}, \citenamefont {Zhong},
  \citenamefont {Gu}, \citenamefont {Yang}, \citenamefont {Lin},\ and\
  \citenamefont {Wen}}]{Du2017}%
  \BibitemOpen
  \bibfield  {author} {\bibinfo {author} {\bibfnamefont {G.}~\bibnamefont
  {Du}}, \bibinfo {author} {\bibfnamefont {Y.}~\bibnamefont {Li}}, \bibinfo
  {author} {\bibfnamefont {J.}~\bibnamefont {Schneeloch}}, \bibinfo {author}
  {\bibfnamefont {R.~D.}\ \bibnamefont {Zhong}}, \bibinfo {author}
  {\bibfnamefont {G.}~\bibnamefont {Gu}}, \bibinfo {author} {\bibfnamefont
  {H.}~\bibnamefont {Yang}}, \bibinfo {author} {\bibfnamefont {H.}~\bibnamefont
  {Lin}}, \ and\ \bibinfo {author} {\bibfnamefont {H.-H.}\ \bibnamefont
  {Wen}},\ }\href {https://doi.org/10.1007/s11433-016-0499-x} {\bibfield
  {journal} {\bibinfo  {journal} {Science China Physics, Mechanics \&
  Astronomy}\ }\textbf {\bibinfo {volume} {60}},\ \bibinfo {pages} {037411}
  (\bibinfo {year} {2017})}\BibitemShut {NoStop}%
\bibitem [{\citenamefont {Pan}\ \emph {et~al.}(2016)\citenamefont {Pan},
  \citenamefont {Nikitin}, \citenamefont {Araizi}, \citenamefont {Huang},
  \citenamefont {Matsushita}, \citenamefont {Naka},\ and\ \citenamefont
  {de~Visser}}]{Pan2016}%
  \BibitemOpen
  \bibfield  {author} {\bibinfo {author} {\bibfnamefont {Y.}~\bibnamefont
  {Pan}}, \bibinfo {author} {\bibfnamefont {A.~M.}\ \bibnamefont {Nikitin}},
  \bibinfo {author} {\bibfnamefont {G.~K.}\ \bibnamefont {Araizi}}, \bibinfo
  {author} {\bibfnamefont {Y.~K.}\ \bibnamefont {Huang}}, \bibinfo {author}
  {\bibfnamefont {Y.}~\bibnamefont {Matsushita}}, \bibinfo {author}
  {\bibfnamefont {T.}~\bibnamefont {Naka}}, \ and\ \bibinfo {author}
  {\bibfnamefont {A.}~\bibnamefont {de~Visser}},\ }\href
  {https://doi.org/10.1038/srep28632} {\bibfield  {journal} {\bibinfo
  {journal} {Scientific Reports}\ }\textbf {\bibinfo {volume} {6}},\ \bibinfo
  {pages} {28632} (\bibinfo {year} {2016})}\BibitemShut {NoStop}%
\bibitem [{\citenamefont {Kuntsevich}\ \emph {et~al.}(2018)\citenamefont
  {Kuntsevich}, \citenamefont {Bryzgalov}, \citenamefont {Prudkoglyad},
  \citenamefont {Martovitskii}, \citenamefont {Selivanov},\ and\ \citenamefont
  {Chizhevskii}}]{Kuntsevich2018}%
  \BibitemOpen
  \bibfield  {author} {\bibinfo {author} {\bibfnamefont {A.~Y.}\ \bibnamefont
  {Kuntsevich}}, \bibinfo {author} {\bibfnamefont {M.~A.}\ \bibnamefont
  {Bryzgalov}}, \bibinfo {author} {\bibfnamefont {V.~A.}\ \bibnamefont
  {Prudkoglyad}}, \bibinfo {author} {\bibfnamefont {V.~P.}\ \bibnamefont
  {Martovitskii}}, \bibinfo {author} {\bibfnamefont {Y.~G.}\ \bibnamefont
  {Selivanov}}, \ and\ \bibinfo {author} {\bibfnamefont {E.~G.}\ \bibnamefont
  {Chizhevskii}},\ }\href {http://dx.doi.org/10.1088/1367-2630/aae595}
  {\bibfield  {journal} {\bibinfo  {journal} {New Journal of Physics}\ }\textbf
  {\bibinfo {volume} {20}},\ \bibinfo {pages} {103022} (\bibinfo {year}
  {2018})}\BibitemShut {NoStop}%
\bibitem [{\citenamefont {Tao}\ \emph {et~al.}(2018)\citenamefont {Tao},
  \citenamefont {Yan}, \citenamefont {Liu}, \citenamefont {Wang}, \citenamefont
  {Ando}, \citenamefont {Wang}, \citenamefont {Zhang},\ and\ \citenamefont
  {Feng}}]{Tao2018}%
  \BibitemOpen
  \bibfield  {author} {\bibinfo {author} {\bibfnamefont {R.}~\bibnamefont
  {Tao}}, \bibinfo {author} {\bibfnamefont {Y.-J.}\ \bibnamefont {Yan}},
  \bibinfo {author} {\bibfnamefont {X.}~\bibnamefont {Liu}}, \bibinfo {author}
  {\bibfnamefont {Z.-W.}\ \bibnamefont {Wang}}, \bibinfo {author}
  {\bibfnamefont {Y.}~\bibnamefont {Ando}}, \bibinfo {author} {\bibfnamefont
  {Q.-H.}\ \bibnamefont {Wang}}, \bibinfo {author} {\bibfnamefont
  {T.}~\bibnamefont {Zhang}}, \ and\ \bibinfo {author} {\bibfnamefont {D.-L.}\
  \bibnamefont {Feng}},\ }\href
  {https://link.aps.org/doi/10.1103/PhysRevX.8.041024} {\bibfield  {journal}
  {\bibinfo  {journal} {Phys. Rev. X}\ }\textbf {\bibinfo {volume} {8}},\
  \bibinfo {pages} {041024} (\bibinfo {year} {2018})}\BibitemShut {NoStop}%
\bibitem [{\citenamefont {Yonezawa}\ \emph {et~al.}(2016)\citenamefont
  {Yonezawa}, \citenamefont {Tajiri}, \citenamefont {Nakata}, \citenamefont
  {Nagai}, \citenamefont {Wang}, \citenamefont {Segawa}, \citenamefont {Ando},\
  and\ \citenamefont {Maeno}}]{Yonezawa2016}%
  \BibitemOpen
  \bibfield  {author} {\bibinfo {author} {\bibfnamefont {S.}~\bibnamefont
  {Yonezawa}}, \bibinfo {author} {\bibfnamefont {K.}~\bibnamefont {Tajiri}},
  \bibinfo {author} {\bibfnamefont {S.}~\bibnamefont {Nakata}}, \bibinfo
  {author} {\bibfnamefont {Y.}~\bibnamefont {Nagai}}, \bibinfo {author}
  {\bibfnamefont {Z.}~\bibnamefont {Wang}}, \bibinfo {author} {\bibfnamefont
  {K.}~\bibnamefont {Segawa}}, \bibinfo {author} {\bibfnamefont
  {Y.}~\bibnamefont {Ando}}, \ and\ \bibinfo {author} {\bibfnamefont
  {Y.}~\bibnamefont {Maeno}},\ }\href {https://doi.org/10.1038/nphys3907}
  {\bibfield  {journal} {\bibinfo  {journal} {Nature Physics}\ }\textbf
  {\bibinfo {volume} {13}},\ \bibinfo {pages} {123} (\bibinfo {year}
  {2016})}\BibitemShut {NoStop}%
\bibitem [{\citenamefont {Willa}\ \emph {et~al.}(2018)\citenamefont {Willa},
  \citenamefont {Willa}, \citenamefont {Song}, \citenamefont {Gu},
  \citenamefont {Schneeloch}, \citenamefont {Zhong}, \citenamefont {Koshelev},
  \citenamefont {Kwok},\ and\ \citenamefont {Welp}}]{Willa2018}%
  \BibitemOpen
  \bibfield  {author} {\bibinfo {author} {\bibfnamefont {K.}~\bibnamefont
  {Willa}}, \bibinfo {author} {\bibfnamefont {R.}~\bibnamefont {Willa}},
  \bibinfo {author} {\bibfnamefont {K.~W.}\ \bibnamefont {Song}}, \bibinfo
  {author} {\bibfnamefont {G.~D.}\ \bibnamefont {Gu}}, \bibinfo {author}
  {\bibfnamefont {J.~A.}\ \bibnamefont {Schneeloch}}, \bibinfo {author}
  {\bibfnamefont {R.}~\bibnamefont {Zhong}}, \bibinfo {author} {\bibfnamefont
  {A.~E.}\ \bibnamefont {Koshelev}}, \bibinfo {author} {\bibfnamefont {W.-K.}\
  \bibnamefont {Kwok}}, \ and\ \bibinfo {author} {\bibfnamefont
  {U.}~\bibnamefont {Welp}},\ }\href
  {https://link.aps.org/doi/10.1103/PhysRevB.98.184509} {\bibfield  {journal}
  {\bibinfo  {journal} {Phys. Rev. B}\ }\textbf {\bibinfo {volume} {98}},\
  \bibinfo {pages} {184509} (\bibinfo {year} {2018})}\BibitemShut {NoStop}%
\bibitem [{\citenamefont {Asaba}\ \emph {et~al.}(2017)\citenamefont {Asaba},
  \citenamefont {Lawson}, \citenamefont {Tinsman}, \citenamefont {Chen},
  \citenamefont {Corbae}, \citenamefont {Li}, \citenamefont {Qiu},
  \citenamefont {Hor}, \citenamefont {Fu},\ and\ \citenamefont
  {Li}}]{Asaba2017}%
  \BibitemOpen
  \bibfield  {author} {\bibinfo {author} {\bibfnamefont {T.}~\bibnamefont
  {Asaba}}, \bibinfo {author} {\bibfnamefont {B.~J.}\ \bibnamefont {Lawson}},
  \bibinfo {author} {\bibfnamefont {C.}~\bibnamefont {Tinsman}}, \bibinfo
  {author} {\bibfnamefont {L.}~\bibnamefont {Chen}}, \bibinfo {author}
  {\bibfnamefont {P.}~\bibnamefont {Corbae}}, \bibinfo {author} {\bibfnamefont
  {G.}~\bibnamefont {Li}}, \bibinfo {author} {\bibfnamefont {Y.}~\bibnamefont
  {Qiu}}, \bibinfo {author} {\bibfnamefont {Y.~S.}\ \bibnamefont {Hor}},
  \bibinfo {author} {\bibfnamefont {L.}~\bibnamefont {Fu}}, \ and\ \bibinfo
  {author} {\bibfnamefont {L.}~\bibnamefont {Li}},\ }\href
  {https://link.aps.org/doi/10.1103/PhysRevX.7.011009} {\bibfield  {journal}
  {\bibinfo  {journal} {Phys. Rev. X}\ }\textbf {\bibinfo {volume} {7}},\
  \bibinfo {pages} {011009} (\bibinfo {year} {2017})}\BibitemShut {NoStop}%
\bibitem [{\citenamefont {Smylie}\ \emph {et~al.}(2018)\citenamefont {Smylie},
  \citenamefont {Willa}, \citenamefont {Claus}, \citenamefont {Koshelev},
  \citenamefont {Song}, \citenamefont {Kwok}, \citenamefont {Islam},
  \citenamefont {Gu}, \citenamefont {Schneeloch}, \citenamefont {Zhong},\ and\
  \citenamefont {Welp}}]{Smylie2018}%
  \BibitemOpen
  \bibfield  {author} {\bibinfo {author} {\bibfnamefont {M.~P.}\ \bibnamefont
  {Smylie}}, \bibinfo {author} {\bibfnamefont {K.}~\bibnamefont {Willa}},
  \bibinfo {author} {\bibfnamefont {H.}~\bibnamefont {Claus}}, \bibinfo
  {author} {\bibfnamefont {A.~E.}\ \bibnamefont {Koshelev}}, \bibinfo {author}
  {\bibfnamefont {K.~W.}\ \bibnamefont {Song}}, \bibinfo {author}
  {\bibfnamefont {W.-K.}\ \bibnamefont {Kwok}}, \bibinfo {author}
  {\bibfnamefont {Z.}~\bibnamefont {Islam}}, \bibinfo {author} {\bibfnamefont
  {G.~D.}\ \bibnamefont {Gu}}, \bibinfo {author} {\bibfnamefont {J.~A.}\
  \bibnamefont {Schneeloch}}, \bibinfo {author} {\bibfnamefont {R.~D.}\
  \bibnamefont {Zhong}}, \ and\ \bibinfo {author} {\bibfnamefont
  {U.}~\bibnamefont {Welp}},\ }\href
  {https://doi.org/10.1038/s41598-018-26032-0} {\bibfield  {journal} {\bibinfo
  {journal} {Scientific Reports}\ }\textbf {\bibinfo {volume} {8}},\ \bibinfo
  {pages} {7666} (\bibinfo {year} {2018})}\BibitemShut {NoStop}%
\bibitem [{\citenamefont {Andersen}\ \emph {et~al.}(2018)\citenamefont
  {Andersen}, \citenamefont {Wang}, \citenamefont {Lorenz},\ and\ \citenamefont
  {Ando}}]{Andersen2018}%
  \BibitemOpen
  \bibfield  {author} {\bibinfo {author} {\bibfnamefont {L.}~\bibnamefont
  {Andersen}}, \bibinfo {author} {\bibfnamefont {Z.}~\bibnamefont {Wang}},
  \bibinfo {author} {\bibfnamefont {T.}~\bibnamefont {Lorenz}}, \ and\ \bibinfo
  {author} {\bibfnamefont {Y.}~\bibnamefont {Ando}},\ }\href
  {https://link.aps.org/doi/10.1103/PhysRevB.98.220512} {\bibfield  {journal}
  {\bibinfo  {journal} {Phys. Rev. B}\ }\textbf {\bibinfo {volume} {98}},\
  \bibinfo {pages} {220512(R)} (\bibinfo {year} {2018})}\BibitemShut {NoStop}%
\bibitem [{\citenamefont {Chen}\ \emph {et~al.}(2018)\citenamefont {Chen},
  \citenamefont {Chen}, \citenamefont {Yang}, \citenamefont {Du},\ and\
  \citenamefont {Wen}}]{Chen2018}%
  \BibitemOpen
  \bibfield  {author} {\bibinfo {author} {\bibfnamefont {M.}~\bibnamefont
  {Chen}}, \bibinfo {author} {\bibfnamefont {X.}~\bibnamefont {Chen}}, \bibinfo
  {author} {\bibfnamefont {H.}~\bibnamefont {Yang}}, \bibinfo {author}
  {\bibfnamefont {Z.}~\bibnamefont {Du}}, \ and\ \bibinfo {author}
  {\bibfnamefont {H.-H.}\ \bibnamefont {Wen}},\ }\href
  {http://advances.sciencemag.org/content/4/6/eaat1084.abstract} {\bibfield
  {journal} {\bibinfo  {journal} {Sci Adv}\ }\textbf {\bibinfo {volume} {4}},\
  \bibinfo {pages} {eaat1084} (\bibinfo {year} {2018})}\BibitemShut {NoStop}%
\bibitem [{\citenamefont {Cho}\ \emph {et~al.}(2019)\citenamefont {Cho},
  \citenamefont {Shen}, \citenamefont {Lyu}, \citenamefont {Lee}, \citenamefont
  {Hor}, \citenamefont {Hecker}, \citenamefont {Schmalian},\ and\ \citenamefont
  {Lortz}}]{Cho2019}%
  \BibitemOpen
  \bibfield  {author} {\bibinfo {author} {\bibfnamefont {C.-w.}\ \bibnamefont
  {Cho}}, \bibinfo {author} {\bibfnamefont {J.}~\bibnamefont {Shen}}, \bibinfo
  {author} {\bibfnamefont {J.}~\bibnamefont {Lyu}}, \bibinfo {author}
  {\bibfnamefont {S.~H.}\ \bibnamefont {Lee}}, \bibinfo {author} {\bibfnamefont
  {Y.~S.}\ \bibnamefont {Hor}}, \bibinfo {author} {\bibfnamefont
  {M.}~\bibnamefont {Hecker}}, \bibinfo {author} {\bibfnamefont
  {J.}~\bibnamefont {Schmalian}}, \ and\ \bibinfo {author} {\bibfnamefont
  {R.}~\bibnamefont {Lortz}},\ }\href {https://arxiv.org/abs/1905.01702}
  {\bibfield  {journal} {\bibinfo  {journal} {arXiv:1905.01702}\ } (\bibinfo
  {year} {2019})}\BibitemShut {NoStop}%
\bibitem [{\citenamefont {Sun}\ \emph {et~al.}(2019)\citenamefont {Sun},
  \citenamefont {Kittaka}, \citenamefont {Sakakibara}, \citenamefont {Machida},
  \citenamefont {Wang}, \citenamefont {Wen}, \citenamefont {Xing},
  \citenamefont {Shi},\ and\ \citenamefont {Tamegai}}]{Sun2019}%
  \BibitemOpen
  \bibfield  {author} {\bibinfo {author} {\bibfnamefont {Y.}~\bibnamefont
  {Sun}}, \bibinfo {author} {\bibfnamefont {S.}~\bibnamefont {Kittaka}},
  \bibinfo {author} {\bibfnamefont {T.}~\bibnamefont {Sakakibara}}, \bibinfo
  {author} {\bibfnamefont {K.}~\bibnamefont {Machida}}, \bibinfo {author}
  {\bibfnamefont {J.}~\bibnamefont {Wang}}, \bibinfo {author} {\bibfnamefont
  {J.}~\bibnamefont {Wen}}, \bibinfo {author} {\bibfnamefont {X.-Z.}\
  \bibnamefont {Xing}}, \bibinfo {author} {\bibfnamefont {Z.}~\bibnamefont
  {Shi}}, \ and\ \bibinfo {author} {\bibfnamefont {T.}~\bibnamefont
  {Tamegai}},\ }\href
  {https://journals.aps.org/prl/abstract/10.1103/PhysRevLett.123.027002}
  {\bibfield  {journal} {\bibinfo  {journal} {Phys. Rev. Lett. 123}\ ,\
  \bibinfo {pages} {027002}} (\bibinfo {year} {2019})}\BibitemShut {NoStop}%
\bibitem [{\citenamefont {Fu}(2014)}]{Fu2014}%
  \BibitemOpen
  \bibfield  {author} {\bibinfo {author} {\bibfnamefont {L.}~\bibnamefont
  {Fu}},\ }\href {https://link.aps.org/doi/10.1103/PhysRevB.90.100509}
  {\bibfield  {journal} {\bibinfo  {journal} {Phys. Rev. B}\ }\textbf {\bibinfo
  {volume} {90}},\ \bibinfo {pages} {100509(R)} (\bibinfo {year}
  {2014})}\BibitemShut {NoStop}%
\bibitem [{\citenamefont {Yonezawa}(2019)}]{Yonezawa2018}%
  \BibitemOpen
  \bibfield  {author} {\bibinfo {author} {\bibfnamefont {S.}~\bibnamefont
  {Yonezawa}},\ }\href@noop {} {\bibfield  {journal} {\bibinfo  {journal}
  {Condensed Matter}\ }\textbf {\bibinfo {volume} {4}},\ \bibinfo {pages} {2}
  (\bibinfo {year} {2019})}\BibitemShut {NoStop}%
\bibitem [{\citenamefont {Li}\ and\ \citenamefont {Xu}(2019)}]{Li2019}%
  \BibitemOpen
  \bibfield  {author} {\bibinfo {author} {\bibfnamefont {Y.}~\bibnamefont
  {Li}}\ and\ \bibinfo {author} {\bibfnamefont {Z.-A.}\ \bibnamefont {Xu}},\
  }\href {\doibase 10.1002/qute.201800112} {\bibfield  {journal} {\bibinfo
  {journal} {Advanced Quantum Technologies}\ }\textbf {\bibinfo {volume} {2}},\
  \bibinfo {pages} {1800112} (\bibinfo {year} {2019})}\BibitemShut {NoStop}%
\bibitem [{\citenamefont {Le}\ \emph {et~al.}(2019)\citenamefont {Le},
  \citenamefont {Sun}, \citenamefont {Jin}, \citenamefont {Che}, \citenamefont
  {Yin}, \citenamefont {Li}, \citenamefont {Pang}, \citenamefont {Xu},
  \citenamefont {Zhao}, \citenamefont {Kittaka}, \citenamefont {Sakakibara},
  \citenamefont {Machida}, \citenamefont {Sankar}, \citenamefont {Yuan},
  \citenamefont {Chen}, \citenamefont {Xu}, \citenamefont {Li}, \citenamefont
  {Zhou},\ and\ \citenamefont {Lu}}]{le2019evidence}%
  \BibitemOpen
  \bibfield  {author} {\bibinfo {author} {\bibfnamefont {T.}~\bibnamefont
  {Le}}, \bibinfo {author} {\bibfnamefont {Y.}~\bibnamefont {Sun}}, \bibinfo
  {author} {\bibfnamefont {H.-K.}\ \bibnamefont {Jin}}, \bibinfo {author}
  {\bibfnamefont {L.}~\bibnamefont {Che}}, \bibinfo {author} {\bibfnamefont
  {L.}~\bibnamefont {Yin}}, \bibinfo {author} {\bibfnamefont {J.}~\bibnamefont
  {Li}}, \bibinfo {author} {\bibfnamefont {G.~M.}\ \bibnamefont {Pang}},
  \bibinfo {author} {\bibfnamefont {C.~Q.}\ \bibnamefont {Xu}}, \bibinfo
  {author} {\bibfnamefont {L.~X.}\ \bibnamefont {Zhao}}, \bibinfo {author}
  {\bibfnamefont {S.}~\bibnamefont {Kittaka}}, \bibinfo {author} {\bibfnamefont
  {T.}~\bibnamefont {Sakakibara}}, \bibinfo {author} {\bibfnamefont
  {K.}~\bibnamefont {Machida}}, \bibinfo {author} {\bibfnamefont
  {R.}~\bibnamefont {Sankar}}, \bibinfo {author} {\bibfnamefont {H.~Q.}\
  \bibnamefont {Yuan}}, \bibinfo {author} {\bibfnamefont {G.~F.}\ \bibnamefont
  {Chen}}, \bibinfo {author} {\bibfnamefont {X.}~\bibnamefont {Xu}}, \bibinfo
  {author} {\bibfnamefont {S.}~\bibnamefont {Li}}, \bibinfo {author}
  {\bibfnamefont {Y.}~\bibnamefont {Zhou}}, \ and\ \bibinfo {author}
  {\bibfnamefont {X.}~\bibnamefont {Lu}},\ }\href@noop {} {\  (\bibinfo {year}
  {2019})},\ \Eprint {http://arxiv.org/abs/1905.11177} {arXiv:1905.11177
  [cond-mat.supr-con]} \BibitemShut {NoStop}%
\bibitem [{\citenamefont {Kriener}\ \emph {et~al.}(2011)\citenamefont
  {Kriener}, \citenamefont {Segawa}, \citenamefont {Ren}, \citenamefont
  {Sasaki},\ and\ \citenamefont {Ando}}]{Kriener2011}%
  \BibitemOpen
  \bibfield  {author} {\bibinfo {author} {\bibfnamefont {M.}~\bibnamefont
  {Kriener}}, \bibinfo {author} {\bibfnamefont {K.}~\bibnamefont {Segawa}},
  \bibinfo {author} {\bibfnamefont {Z.}~\bibnamefont {Ren}}, \bibinfo {author}
  {\bibfnamefont {S.}~\bibnamefont {Sasaki}}, \ and\ \bibinfo {author}
  {\bibfnamefont {Y.}~\bibnamefont {Ando}},\ }\href
  {https://link.aps.org/doi/10.1103/PhysRevLett.106.127004} {\bibfield
  {journal} {\bibinfo  {journal} {Phys. Rev. Lett.}\ }\textbf {\bibinfo
  {volume} {106}},\ \bibinfo {pages} {127004} (\bibinfo {year}
  {2011})}\BibitemShut {NoStop}%
\bibitem [{\citenamefont {Venderbos}\ \emph {et~al.}(2016)\citenamefont
  {Venderbos}, \citenamefont {Kozii},\ and\ \citenamefont
  {Fu}}]{Venderbos2016}%
  \BibitemOpen
  \bibfield  {author} {\bibinfo {author} {\bibfnamefont {J.~W.~F.}\
  \bibnamefont {Venderbos}}, \bibinfo {author} {\bibfnamefont {V.}~\bibnamefont
  {Kozii}}, \ and\ \bibinfo {author} {\bibfnamefont {L.}~\bibnamefont {Fu}},\
  }\href {https://link.aps.org/doi/10.1103/PhysRevB.94.094522} {\bibfield
  {journal} {\bibinfo  {journal} {Phys. Rev. B}\ }\textbf {\bibinfo {volume}
  {94}},\ \bibinfo {pages} {094522} (\bibinfo {year} {2016})}\BibitemShut
  {NoStop}%
\bibitem [{\citenamefont {Kuntsevich}\ \emph {et~al.}(2019)\citenamefont
  {Kuntsevich}, , \citenamefont {Martovitskii}, \citenamefont {Rybalchenko},
  \citenamefont {Selivanov}, \citenamefont {Sobolevskiy},\ and\ \citenamefont
  {Chizhevskii}}]{Kuntsevich2019}%
  \BibitemOpen
  \bibfield  {author} {\bibinfo {author} {\bibfnamefont {A.~Y.}\ \bibnamefont
  {Kuntsevich}}, , \bibinfo {author} {\bibfnamefont {V.~P.}\ \bibnamefont
  {Martovitskii}}, \bibinfo {author} {\bibfnamefont {G.~V.}\ \bibnamefont
  {Rybalchenko}}, \bibinfo {author} {\bibfnamefont {Y.~G.}\ \bibnamefont
  {Selivanov}}, \bibinfo {author} {\bibfnamefont {O.~A.}\ \bibnamefont
  {Sobolevskiy}}, \ and\ \bibinfo {author} {\bibfnamefont {E.~G.}\ \bibnamefont
  {Chizhevskii}},\ }\href@noop {} {\bibfield  {journal} {\bibinfo  {journal}
  {Materials}\ }\textbf {\bibinfo {volume} {12}},\ \bibinfo {pages} {37XX}
  (\bibinfo {year} {2019})}\BibitemShut {NoStop}%
\bibitem [{\citenamefont {Liu}\ \emph {et~al.}(2015)\citenamefont {Liu},
  \citenamefont {Yao}, \citenamefont {Shao}, \citenamefont {Ming},
  \citenamefont {Pi}, \citenamefont {Tan}, \citenamefont {Zhang},\ and\
  \citenamefont {Zhang}}]{Liu2015}%
  \BibitemOpen
  \bibfield  {author} {\bibinfo {author} {\bibfnamefont {Z.}~\bibnamefont
  {Liu}}, \bibinfo {author} {\bibfnamefont {X.}~\bibnamefont {Yao}}, \bibinfo
  {author} {\bibfnamefont {J.}~\bibnamefont {Shao}}, \bibinfo {author}
  {\bibfnamefont {Z.}~\bibnamefont {Ming}}, \bibinfo {author} {\bibfnamefont
  {L.}~\bibnamefont {Pi}}, \bibinfo {author} {\bibfnamefont {S.}~\bibnamefont
  {Tan}}, \bibinfo {author} {\bibfnamefont {C.}~\bibnamefont {Zhang}}, \ and\
  \bibinfo {author} {\bibfnamefont {Y.}~\bibnamefont {Zhang}},\ }\href
  {https://pubs.acs.org/doi/10.1021/jacs.5b06815} {\bibfield  {journal}
  {\bibinfo  {journal} {Journ. Amer. Chem. Soc}\ }\textbf {\bibinfo {volume}
  {137}},\ \bibinfo {pages} {10512} (\bibinfo {year} {2015})}\BibitemShut
  {NoStop}%
\bibitem{Supl} {Supplemental Material can be found at [TO ADD URL]. It contains theories of nematic superconductivity and normal state magnetorestance, procedure of sample selection, description of hexagonal notations, summary of structural and transport data for 9 samples, and explanation why we indeed selected single blocks.}
\bibitem{Bond}{W.L. Bond, Crystal Technology, New York: Wiley, 1976.}
\bibitem{Sungawa} {I. Sunagawa, Crystals. Growth, Morphology and Perfection. Cambridge University Press, 2005.}
\bibitem{Volosheniuk2019} {S.O. Volosheniuk,  Yu.G. Selivanov, M.A. Bryzgalov, V. P. Martovitskii,  A. Yu. Kuntsevich, Journ. of Appl. Phys. {\bf 125}, 095103 (2019).}
\bibitem{fu2009} {L. Fu, %Hexagonal Warping Effects in the Surface States of the Topological Insulator $Bi_2Te_3$, 
\prl {\bf 103} 266801 (2009).}
\bibitem{Proskurin}{I. Proskurin, M. Ogata, Y. Suzumura, \prb {\bf 91}, 195413 (2015).}
\bibitem [{\citenamefont {Akzyanov}\ and\ \citenamefont
  {Rakhmanov}(2018)}]{Akzyanov2018}%
  \BibitemOpen
  \bibfield  {author} {\bibinfo {author} {\bibfnamefont {R.~S.}\ \bibnamefont
  {Akzyanov}}\ and\ \bibinfo {author} {\bibfnamefont {A.~L.}\ \bibnamefont
  {Rakhmanov}},\ }\href {https://link.aps.org/doi/10.1103/PhysRevB.97.075421}
  {\bibfield  {journal} {\bibinfo  {journal} {Phys. Rev. B}\ }\textbf {\bibinfo
  {volume} {97}},\ \bibinfo {pages} {075421} (\bibinfo {year}
  {2018})}\BibitemShut {NoStop}%
\bibitem [{\citenamefont {Chiba}\ \emph {et~al.}(2017)\citenamefont {Chiba},
  \citenamefont {Takahashi},\ and\ \citenamefont {Bauer}}]{Chiba2017}%
  \BibitemOpen
  \bibfield  {author} {\bibinfo {author} {\bibfnamefont {T.}~\bibnamefont
  {Chiba}}, \bibinfo {author} {\bibfnamefont {S.}~\bibnamefont {Takahashi}}, \
  and\ \bibinfo {author} {\bibfnamefont {G.~E.~W.}\ \bibnamefont {Bauer}},\
  }\href {https://link.aps.org/doi/10.1103/PhysRevB.95.094428} {\bibfield
  {journal} {\bibinfo  {journal} {Phys. Rev. B}\ }\textbf {\bibinfo {volume}
  {95}},\ \bibinfo {pages} {094428} (\bibinfo {year} {2017})}\BibitemShut
  {NoStop}%
\bibitem [{\citenamefont {Taskin}\ \emph {et~al.}(2017)\citenamefont {Taskin},
  \citenamefont {Legg}, \citenamefont {Yang}, \citenamefont {Sasaki},
  \citenamefont {Kanai}, \citenamefont {Matsumoto}, \citenamefont {Rosch},\
  and\ \citenamefont {Ando}}]{Taskin2017}%
  \BibitemOpen
  \bibfield  {author} {\bibinfo {author} {\bibfnamefont {A.}~\bibnamefont
  {Taskin}}, \bibinfo {author} {\bibfnamefont {H.~F.}\ \bibnamefont {Legg}},
  \bibinfo {author} {\bibfnamefont {F.}~\bibnamefont {Yang}}, \bibinfo {author}
  {\bibfnamefont {S.}~\bibnamefont {Sasaki}}, \bibinfo {author} {\bibfnamefont
  {Y.}~\bibnamefont {Kanai}}, \bibinfo {author} {\bibfnamefont
  {K.}~\bibnamefont {Matsumoto}}, \bibinfo {author} {\bibfnamefont
  {A.}~\bibnamefont {Rosch}}, \ and\ \bibinfo {author} {\bibfnamefont
  {Y.}~\bibnamefont {Ando}},\ }\href
  {https://www.nature.com/articles/s41467-017-01474-8} {\bibfield  {journal}
  {\bibinfo  {journal} {Nature Communications}, \textbf{\bibinfo {volume} {8}},\ \bibinfo {pages} {1340}}
  (\bibinfo {year} {2017})}\BibitemShut {NoStop}%
\bibitem [{\citenamefont {Huang}\ \emph {et~al.}(2018)\citenamefont {Huang},
  \citenamefont {Gu}, \citenamefont {Ji}, \citenamefont {Wang}, \citenamefont
  {Hu}, \citenamefont {Qin}, \citenamefont {Wang},\ and\ \citenamefont
  {Zhang}}]{Huang2018}%
  \BibitemOpen
  \bibfield  {author} {\bibinfo {author} {\bibfnamefont {H.}~\bibnamefont
  {Huang}}, \bibinfo {author} {\bibfnamefont {J.}~\bibnamefont {Gu}}, \bibinfo
  {author} {\bibfnamefont {P.}~\bibnamefont {Ji}}, \bibinfo {author}
  {\bibfnamefont {Q.}~\bibnamefont {Wang}}, \bibinfo {author} {\bibfnamefont
  {X.}~\bibnamefont {Hu}}, \bibinfo {author} {\bibfnamefont {Y.}~\bibnamefont
  {Qin}}, \bibinfo {author} {\bibfnamefont {J.}~\bibnamefont {Wang}}, \ and\
  \bibinfo {author} {\bibfnamefont {C.}~\bibnamefont {Zhang}},\ }\bibfield
  {booktitle} {\emph {\bibinfo {booktitle} {Applied Physics Letters}},\ }\href
  {\doibase 10.1063/1.5063689} {\bibfield  {journal} {\bibinfo  {journal}
  {Appl. Phys. Lett.}\ }\textbf {\bibinfo {volume} {113}},\ \bibinfo {pages}
  {222601} (\bibinfo {year} {2018})}\BibitemShut {NoStop}%
\bibitem [{\citenamefont {Hecker}\ and\ \citenamefont
  {Schmalian}(2017)}]{Hecker2018}%
  \BibitemOpen
  \bibfield  {author} {\bibinfo {author} {\bibfnamefont {M.}~\bibnamefont
  {Hecker}}\ and\ \bibinfo {author} {\bibfnamefont {J.}~\bibnamefont
  {Schmalian}},\ }\href {https://www.nature.com/articles/s41535-018-0098-z.pdf}
  {\bibfield  {journal} {\bibinfo  {journal} {npj Quantum Mater.}, \textbf{\bibinfo {volume} {3}}, \bibinfo
  {pages} {26}} (\bibinfo {year} {2017})}\BibitemShut {NoStop}%
\end{thebibliography}
\end{document}